\documentclass[twoside,twocolumn,9pt]{article}
\usepackage{extsizes}
\usepackage[super,sort&compress,comma]{natbib} 
\usepackage[version=3]{mhchem}
\usepackage[left=1.5cm, right=1.5cm, top=1.785cm, bottom=2.0cm]{geometry}
\usepackage{balance}
\usepackage{mathptmx}
\usepackage{sectsty}
\usepackage{graphicx} 
\usepackage{lastpage}
\usepackage[format=plain,justification=justified,singlelinecheck=false,font={stretch=1.125,small,sf},labelfont=bf,labelsep=space]{caption}
\usepackage{float}
\usepackage{fancyhdr}
\usepackage{fnpos}
\usepackage[english]{babel}
\addto{\captionsenglish}{%
  
}
\usepackage{array}
\usepackage{droidsans}
\usepackage{charter}
\usepackage[T1]{fontenc}
\usepackage[usenames,dvipsnames]{xcolor}
\usepackage{setspace}
\usepackage[compact]{titlesec}
\usepackage{hyperref}

\usepackage{epstopdf}

\definecolor{cream}{RGB}{222,217,201}

\begin{document}

\pagestyle{fancy}
\thispagestyle{plain}
\fancypagestyle{plain}{
\renewcommand{\headrulewidth}{0pt}
}

\makeFNbottom
\makeatletter
\renewcommand\LARGE{\@setfontsize\LARGE{15pt}{17}}
\renewcommand\Large{\@setfontsize\Large{12pt}{14}}
\renewcommand\large{\@setfontsize\large{10pt}{12}}
\renewcommand\footnotesize{\@setfontsize\footnotesize{7pt}{10}}
\makeatother

\renewcommand{\thefootnote}{\fnsymbol{footnote}}
\renewcommand\footnoterule{\vspace*{1pt}%
\color{cream}\hrule width 3.5in height 0.4pt \color{black}\vspace*{5pt}} 
\setcounter{secnumdepth}{5}

\makeatletter 
\renewcommand\@biblabel[1]{#1}            
\renewcommand\@makefntext[1]%
{\noindent\makebox[0pt][r]{\@thefnmark\,}#1}
\makeatother 
\renewcommand{\figurename}{\small{Fig.}~}
\sectionfont{\sffamily\Large}
\subsectionfont{\normalsize}
\subsubsectionfont{\bf}
\setstretch{1.125} 
\setlength{\skip\footins}{0.8cm}
\setlength{\footnotesep}{0.25cm}
\setlength{\jot}{10pt}
\titlespacing*{\section}{0pt}{4pt}{4pt}
\titlespacing*{\subsection}{0pt}{15pt}{1pt}

\fancyfoot{}
\fancyfoot[LO,RE]{\vspace{-7.1pt}\includegraphics[height=9pt]{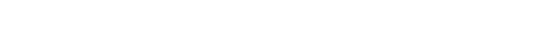}}
\fancyfoot[CO]{\vspace{-7.1pt}\hspace{13.2cm}\includegraphics{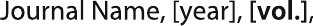}}
\fancyfoot[CE]{\vspace{-7.2pt}\hspace{-14.2cm}\includegraphics{head_foot/RF}}
\fancyfoot[RO]{\footnotesize{\sffamily{1--\pageref{LastPage} ~\textbar  \hspace{2pt}\thepage}}}
\fancyfoot[LE]{\footnotesize{\sffamily{\thepage~\textbar\hspace{3.45cm} 1--\pageref{LastPage}}}}
\fancyhead{}
\renewcommand{\headrulewidth}{0pt} 
\renewcommand{\footrulewidth}{0pt}
\setlength{\arrayrulewidth}{1pt}
\setlength{\columnsep}{6.5mm}
\setlength\bibsep{1pt}

\makeatletter 
\newlength{\figrulesep} 
\setlength{\figrulesep}{0.5\textfloatsep} 

\newcommand{\topfigrule}{\vspace*{-1pt}%
\noindent{\color{cream}\rule[-\figrulesep]{\columnwidth}{1.5pt}} }

\newcommand{\botfigrule}{\vspace*{-2pt}%
\noindent{\color{cream}\rule[\figrulesep]{\columnwidth}{1.5pt}} }

\newcommand{\dblfigrule}{\vspace*{-1pt}%
\noindent{\color{cream}\rule[-\figrulesep]{\textwidth}{1.5pt}} }

\makeatother

\twocolumn[
  \begin{@twocolumnfalse}
{\includegraphics[height=30pt]{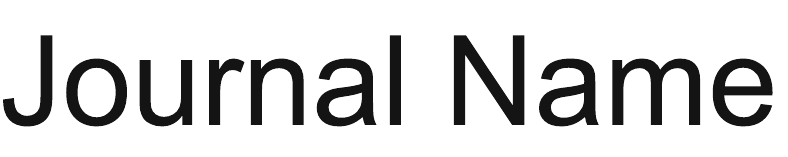}\hfill\raisebox{0pt}[0pt][0pt]{\includegraphics[height=55pt]{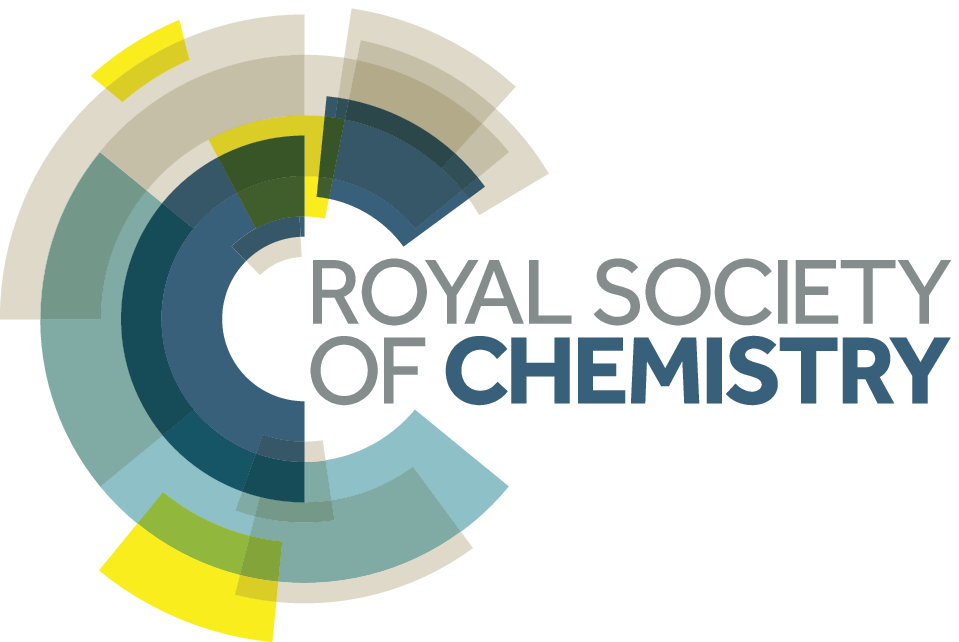}}\\[1ex]
\includegraphics[width=18.5cm]{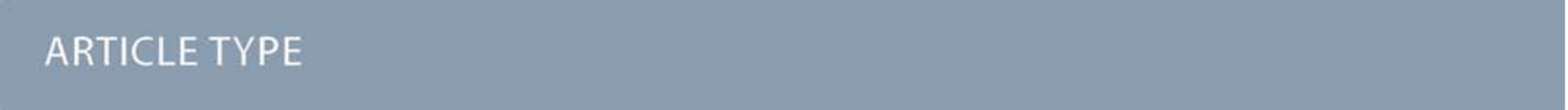}}\par
\vspace{1em}
\sffamily
\begin{tabular}{m{4.5cm} p{13.5cm} }

\includegraphics{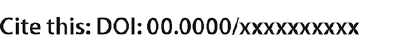} & \noindent\LARGE{\textbf{Feature Fusion of Raman Chemical Imaging and
Digital Histopathology using Machine Learning for Prostate Cancer Detection}} \\
\vspace{0.3cm} & \vspace{0.3cm} \\

 & \noindent\large{Trevor Doherty,\textit{$^{a}$$^{\ast}$} Susan McKeever,\textit{$^{a}$} Nebras Al-Attar,\textit{$^{b,}$}\textit{$^{g}$} Tiarnán Murphy,\textit{$^{b}$} Claudia Aura,\textit{$^{c}$} Arman Rahman,\textit{$^{c}$} Amanda O’Neill,\textit{$^{d}$} Stephen P Finn,\textit{$^{e}$} Elaine Kay,\textit{$^{f}$} William M. Gallagher,\textit{$^{c}$} R. William G. Watson,\textit{$^{d}$}  Aoife Gowen\textit{$^{c}$$^{\ddag}$} \& Patrick Jackman\textit{$^{a}$$^{\ddag}$}} \\

\includegraphics{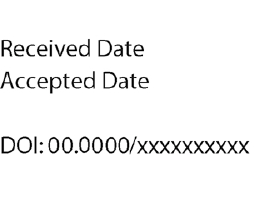} & \noindent\normalsize{The diagnosis of prostate cancer is challenging due to the heterogeneity of its presentations,   leading to the over diagnosis and treatment of non-clinically important disease. Accurate diagnosis can directly benefit a patient’s quality of life and prognosis.  Towards addressing this issue, we present a learning model for the automatic identification of prostate cancer. While many prostate cancer studies have adopted Raman spectroscopy approaches, none have utilised the combination of Raman Chemical Imaging (RCI) and other imaging modalities. This study uses multimodal images formed from stained Digital Histopathology (DP) and unstained RCI. 
The approach was developed and tested on a set of 178 clinical samples from 32 patients, containing a range of non-cancerous, Gleason grade 3 (G3) and grade 4 (G4) tissue microarray samples. For each histological sample, there is a pathologist labelled DP - RCI image pair. The hypothesis tested was whether multimodal image models can outperform single modality baseline models in terms of diagnostic accuracy. 
Binary non-cancer/cancer models and the more challenging G3/G4 differentiation were investigated. Regarding G3/G4 classification, the multimodal approach achieved a sensitivity of 73.8\% and specificity of 88.1\% while the baseline DP model showed a sensitivity and specificity of 54.1\% and 84.7\% respectively. The multimodal approach demonstrated a statistically significant 12.7\% AUC advantage over the baseline with a value of 85.8\% compared to 73.1\%, also outperforming models based solely on RCI and median Raman spectra. Feature fusion of DP and RCI does not improve the more trivial task of tumour identification but does deliver an observed advantage in G3/G4 discrimination. Building on these promising findings, future work could include the acquisition of larger datasets for enhanced model generalization.  
} \\

\end{tabular}

 \end{@twocolumnfalse} \vspace{0.4cm}

  ]

\renewcommand*\rmdefault{bch}\normalfont\upshape
\rmfamily
\section*{}
\vspace{-1cm}


\footnotetext{$^{\ast}$\textit{~Corresponding author}}
\footnotetext{\textit{$^{a}$~Technological University Dublin, City Campus, Grangegorman Lower, Dublin 7, Ireland.}}
\footnotetext{\textit{$^{b}$~University College Dublin, School of Biosystems and Food Engineering, Belfield, Dublin 4, Ireland.}}
\footnotetext{\textit{$^{c}$~University College Dublin, UCD School of Biomolecular and Biomedical Science, UCD Conway Institute, Belfield, Dublin 4, Ireland.}}
\footnotetext{\textit{$^{d}$~University College Dublin, UCD School of Medicine, UCD Conway Institute, Belfield, Dublin 4, Ireland.}}
\footnotetext{\textit{$^{e}$~Departments of Histopathology and Cancer Molecular Diagnostics, St. James Hospital and Trinity College Dublin.}}
\footnotetext{\textit{$^{f}$~Department of Pathology, Beaumont Hospital, Beaumont Rd, Beaumont, Dublin 9, Ireland.}}
\footnotetext{\textit{$^{g}$~Department of Laser and Optoelectronics Engineering, University of Technology, Baghdad, Iraq.}}


\footnotetext{\ddag \textit~These authors share last authorship.}


\section{Introduction}
Cancer is a primary cause of morbidity and mortality worldwide with prostate cancer having the second highest incidence of cancer among men.\cite{stewart20162014} It is estimated that prostate cancer will account for 33,330 deaths and 191,930 new cases in the U.S. in 2020.\cite{siegel2020cancer} Manual histologic assessment of prostate tissue biopsies by a pathologist constitutes the definitive diagnosis of prostate cancer.\cite{humphrey2003prostate}
Pathologists utilize advanced imaging techniques and biomarker discovery through the use of Haematoxylin and Eosin (H\&E) to visually analyze the morphological features within a prostate tissue sample and use the Gleason grading system \cite{lomas2020all, gleason1974prediction} to identify nonmalignant, benign hypertrophic and malignant prostatic tissues. Limitations of this staining process including reactivity to staining agents from different manufacturers, tissues’ thickness, stain concentration, batch effects, room temperature during preparation and alteration of the applied method lead to staining variation across use cases.\cite{bentaieb2017adversarial, lyon1994standardization, leo2018stable} This, in addition to factors such as the extent of a pathologist’s skill and training in Gleason grading, weariness due to the laborious assessment procedure and subjectivity introduce error into the pathologist report and have led to high rates of interobserver variability.\cite{allsbrook2001interobserver, borowsky2020digital,mosquera2014computer} Computer-aided diagnosis can be employed to support pathologists, increase efficiency and accuracy and ultimately benefit the patient.\cite{irshad2013methods}\par 
Fundamental differences in the molecular expression of a range of diseases including some cancers tend to present themselves as changes in tissue architecture and nuclear morphology. As a result, there has been significant interest in the use of digital pathology to develop algorithms and feature extraction methods for the automatic classification of tissue types, the grading of disease and prediction of disease prognosis.\cite{madabhushi2016image} In addition to digital pathology, spectroscopic methods are highly suitable for the in situ observation of molecular changes, allowing characterization of complex biological systems.\cite{dybas2016raman} By interrogating chemical content, Raman scattering can produce a detailed fingerprint of a material. \cite{baker2018clinical} 
Raman spectroscopy as the most dominant spectral technique has shown details of molecular information with the ability to provide qualitative and quantitative biochemical signatures in addition to sample morphology.\cite{krafft2017label} Furthermore, Raman spectroscopy can simultaneously identify and detect a variety of molecular structures such as proteins, DNA, RNA, and lipids in a single acquisition.\cite{roman2019exploring} Chemically complex milieux of biological samples revealed by Raman spectroscopy allow investigation of functional groups, bonding types and molecular conformations and have led to markers for cancer detection.\cite{movasaghi2007raman} Consequently, Raman spectroscopy has been applied to a range of cancer modalities such as colon, \cite{chowdary2007discrimination, brozek2019analysis} skin,\cite{gniadecka2004melanoma} breast,\cite{haka2005diagnosing} lung,\cite{huang2003near} cervix\cite{lyng2007vibrational} and prostate\cite{crow2003use, devpura2010detection} among others.\par
Prostate adenocarcinoma has shown to be difficult to accurately characterize, partially due to histological heterogeneity of samples. This characteristic heterogeneity is more suited to an approach based on image analysis as opposed to ’point’ spectroscopy.\cite{tollefson2010raman} Kast et. al suggested that an opportunity exists to combine the power of Raman spectroscopy with imaging technologies as this would allow correlation of sample morphology with molecular information.\cite{kast2014emerging}
Raman chemical imaging (RCI) is a novel emerging approach that utilises conventional Raman spectroscopy with a spatial mapping process that results in a detailed chemical map of the sample containing biochemical information that is imperceptible by conventional methods. Utilising RCI, Samiei et al. found that it showed promise as a technique in the identification of patients who were at risk of post-radical prostatectomy progression, indicating distinctive chemical differences in biochemical failure Gleason 7 patients,\cite{samiei2018prospective} while Tollefson et al. observed distinctive chemical differences between prostate tissue for patients that progressed to metastatic disease and those that didn’t.\cite{tollefson2010raman}\par
While many prostate cancer studies in the current literature utilise Raman spectroscopy-based approaches, \cite{crow2005assessment, li2014noninvasive, wang2013raman, crow2005use, del2015surface} none have explored the combination of RCI with other imaging modalities. RCI applied to the identification of cancer biomarkers in prostate tissue could lead to improvements in current identification processes by complementing traditional staining methods. Image fusion methods incorporating multi-sensor and multi-source data offer a wider diversity of potential features for medical analysis applications.\cite{du2016overview, james2014medical}
Adopting a multimodal microscopy approach for the automated analysis of prostate cancer, Kwak et al. combined imaging data from Fourier transform infra-red and optical microscopy and achieved better accuracy on the fusion of these data sources than a single data source alone.\cite{kwak2011multimodal} The combination of coherent anti-Stokes Raman scattering, two-photon excited fluorescence and second harmonic generation microscopy showed promise in differentiation of head and neck cancers and non-cancerous epithelium \cite{rodner2019fully} while Patil et al. demonstrated that the integration of Raman spectroscopy with Optical Coherence Tomography (OCT) enhanced the detection of ex-vivo breast cancer when compared with either approach alone.\cite{patil2008combined} 
Investigating colonic adenocarcinoma, Ashok et al. used Raman spectroscopy and OCT which individually yielded sensitivities of 89\% and 77\% respectively and specificities of 78\% and 74\% respectively. Both sensitivity and specificity improved to 94\% for the combined imaging model.\cite{ashok2013multi} Yuan et al. devised a predictor for breast cancer survival that combined both image and gene expression analyses that significantly outperformed models that used only a single data source.\cite{yuan2012quantitative} This study proposes the development of a multimodal machine learning model that integrates features from both digital histopathology and RCI.\par
Many approaches extract quantitative morphological information from histopathology sample images. For the automatic Gleason grading of prostate cancer, Nguyen et al. extracted morphological and phase based features relating to glands and surrounding stroma to predict disease pathology.\cite{nguyen2017automatic} Lee et al. utilised domain inspired features for the prediction of biochemical recurrence of prostate cancer \cite{lee2014co} while, in a study on male breast cancer, Veta et al. found mean nuclear area to be a significant prognostic indicator.\cite{veta2012prognostic} Doyle et al. utilized architectural and textural features to achieve an accuracy of 76.9\% for the classification of Gleason grade 3 and 4 histological tissue patches.\cite{doyle2007automated} Novel explicit shape descriptors have been used to differentiate between intermediate Gleason grade (grades 3 and 4) prostate glands using histopathology images with an accuracy of 89\% and AUC of 0.78 being reported.\cite{sparks2013explicit}
Linder et al. utilised local binary patterns for the segmentation of tumour stromal and epithelial regions using bright field microscopy for colorectal cancer.\cite{linder2012identification} Jafari-Khouzani and Soltanian-Zadeh utilized energy and entropy features of multiwavelet coefficients in order to classify prostate cancer Gleason scores from optical microscopy images.\cite{jafari2003multiwavelet} Sanghavi et al. adopted Speeded Up Robust Features (SURF) and Scale Invariant Feature Transform (SIFT) with Bag of Visual Words (BoVW) to identify prostate cancer grades from histopathology images.\cite{sanghavi2016automated} \par 
SIFT is a feature detection algorithm from the field of computer vision which locates and describes distinctive invariant image features.\cite{lowe2004distinctive} Clustering of feature vectors yields cluster centroids that characterize visual words. Pretrained dictionaries can be used to map descriptors within new images to the closest visual word. An image can then be represented as a histogram of its component visual words, generating an input vector for a classification algorithm.
Deep learning and transfer learning have shown efficacy when applied to the analysis and classification of cancer and other diseases.\cite{bychkov2018deep, ertosun2015automated, shie2015transfer, weng2017combining, esteva2017dermatologist, ribeiro2016exploring, chi2017thyroid, shin2016deep} A drawback of deep learning is that it requires large amounts of training data, which is not commonly the case in the medical domain. Several studies have demonstrated promise on prostate cancer detection using deep learning, however these studies have utilised relatively large patient cohorts and image sets. Ishioka,\cite{ishioka2018computer} Arvaniti,\cite{arvaniti2018automated} Litjens \cite{litjens2016deep} and Song’s\cite{song2018computer} studies contained 335, 886, 225 and 195 patients respectively. 
Where large numbers of samples are not available and deep learning is not applicable, well established computer vision techniques such as SIFT come into play.\cite{o2019deep} Wang et al. chose the SIFT image feature with bag-of-word model as the non-deep learning representative method in their comparison of deep-learning versus non-deep learning for the classification of prostate cancer using MRI images.\cite{wang2017searching} SIFT with BoVW was one of the main representative methods for computer aided diagnosis algorithms before the era of deep learning as can be seen in the results of ImageNet 2012.\cite{russakovsky2015imagenet} It is, therefore, still a valuable approach when dataset size precludes the use of deep learning technology and is used in this study.\par
The current study which developed models that classify prostate tissue samples goes beyond existing studies by integrating features of both Raman chemical and histopathology images. The main research question was whether models trained on multimodal images can classify prostate cancer tissue samples more effectively than a range of single modality baseline models utilizing digital histopathology imaging, Raman chemical imaging or median Raman spectra.
Specifically, the study aimed to investigate these approaches through the binary classifications of cancer/non-cancer and Gleason grade 3/ grade 4 tissue samples with the latter case relating to the much more challenging and clinically relevant assessment of cancer severity. This research constitutes an initial evaluation of the approach with more data being necessary for a generalizable model.

\section{Materials and Methods}

\subsection{Histopathology sample preparation Raman Mapping}
A formalin fixed paraffin embedded prostate tissue microarray (TMA) of 32 patients was obtained from the Irish Prostate Cancer Research Consortium (PCRC) with ethical approval and patient written consent. This TMA set consists of three TMA blocks comprising 178 cores. The TMA sections were cut and placed on a standard microscopic slide. The optimal tissue thickness was investigated through employing various block cut thicknesses (4,8,10, and 16 $\mu$m).\cite{breen2017investigating} A 10 $\mu$m TMA thickness was the optimal choice for Raman imaging as it provided spectra that were relatively free of interference from the substrate. Additionally, this thickness allowed a sufficient layer of tissue to include features such as epithelial cells, stroma, fibroblast, and red blood cells.\par
Samples were then baked in the oven at $60^\circ$C for one hour and allowed to dry. A de-paraffinization protocol was introduced to effectively remove paraffin from the tissue sections, described as follows: samples were consecutively dewaxed in three baths of xylene for 6 minutes each followed by three baths of graded ethanol (100\%, 90\%, and 80\%), also for 6 minutes each. The samples were rehydrated by immersing them in deionized water for 6 minutes. The rehydrated tissues were then kept in deionized water in petri dishes in advance of Raman imaging tissue spectral acquisition. Rehydrated samples were imaged on standard glass slides under water immersion. After Raman imaging tissue samples were manually stained with Haematoxylin-Eosin.\par
Raman imaging spectroscopy was performed using a Renishaw inVia Micro-Raman confocal spectroscopy system (Renishaw, Wotton-under-Edge, Gloucestershire, UK) fitted with a 532nm and 785nm solid-state diode laser. The 532nm laser was selected to perform Raman Imaging. Raman spectra were recorded using water immersion x63 objective and a grating (600 line/mm) with a 1 second acquisition time. Before spectral data was gathered a brightfield montage of the entire slide was collected in order to assist in targeting the TMA cores and minimise the amount of data collected from empty sections of the slide. Experiments could then be queued to run consecutively. Multiple single point scans were used to perform map image acquisition. A rectangular grid was created to sample the core surface using WiRE software version 4.1. The dimensions of the grid step size were 33$\mu$m in the x direction and 33$\mu$m in the y direction. The total  mapped area varied according to the slight variability in TMA core size, ranging from 19x31 pixels to 48x37 pixels.

\subsection{TMA digitization}
Hematoxylin solution modified acc. to Gill II and Eosin Y-solution 0.5\% aqueous were purchased from Sigma-Aldrich to perform staining on the PCRC Cohort after completion of Raman acquisition. H\&E staining was performed in one batch using routine clinical protocols.\cite{stmichaels} H\&E slides were cover-slipping and digitized at 20x magnification using a whole slide scanner (Aperio ScanScope CS, Leica Biosystems, Buffalo Grove, IL, United States). For cross-validation of Gleason score (GS), the digitized H\&E TMA images were investigated at the core-level by histopathology. These images were annotated corresponding to the TMA map core description using Oncotopix\cite{rahman2020advances} (from Visiopharm) Version 2019.07 platform. Hence, the TMA map was loaded to the Visiopharm platform to obtain TIFF format images for individual cores which correspond to patient ID. This allows the pathologist to validate the H\&E staining, GS crosscheck, and annotate the tumours for individual cores. Meantime, the cores containing artifacts of the histological processing (e.g., tissue folds, debris), were demarcated using the same tools. These cores were ultimately excluded from Raman Imaging data. Digitized cores and their annotations were stored and managed.

\subsection{Patient Cohort and Sample Details}
To investigate the effect of augmenting digital pathology images with RCI, a study was conducted on a cohort of 32 prostate cancer patients. 178 TMA samples were obtained as multiple samples could be extracted from a single patient’s radical prostatectomy tissue.\cite{breen2017investigating} Cancer TMA samples were pathologically confirmed normal, Gleason grade 3 or 4, as according to the paper by Breen et al.\cite{breen2017investigating} Raman chemical images of the samples were acquired with the pathologically verified label being transferred from the digital pathology sample to its Raman counterpart. Table 1 outlines the data summary and class distribution of non-cancer, Gleason grade 3 and grade 4 samples.

\begin{table}[h]
\small
  \caption{\ Class balance details for 32-patient cohort}
  \label{tbl:example}
  \begin{tabular*}{0.48\textwidth}{m{2cm} m{2cm} m{2cm} m{2cm}}
    \hline
    \textbf{Class} & \textbf{No. of Digital Pathology \newline Images} & \textbf{No. of \newline Raman \newline Chemical \newline Images} & \textbf{\% of Total \newline Samples}\\
    \hline
    Non-cancer & 82 & 82 & 45.8\% \\
    Cancer & 97 & 97 & 54.2\% \\
    \hline
    \textbf{Total} & \textbf{179} & \textbf{179} & \textbf{100.0\%} \\
    \hline
    \textbf{Cancer \newline Breakdown} &  & & \\
    Gleason 3 & 51 & 51 & 52.6\% \\
    Gleason 4 & 45 & 45 & 46.3\% \\
    Gleason 5 & 1 & 1 & 1.0\% \\
    \hline
    \textbf{Total} & \textbf{97} & \textbf{97} & \textbf{100.0\%} \\
    \hline
  \end{tabular*}
\end{table}

\subsection{Image pre-processing}
Pretreatment of Raman spectra included the removal of saturated pixels followed by cosmic ray correction using in-house functions in MATLAB 2019a. Spectra were used to form Raman chemical images with each spectrum constituting a single pixel of a tissue sample image.
Each TMA sample’s Raman chemical image was approximately 33 x 33 pixels. The mean of each pixel’s spectrum was used to form a grey scale 2-D Raman chemical image. The Raman chemical images were then upsampled to 500 x 500 pixels using cubic spline interpolation. Up-sampling better facilitated the discovery of useful local and global features before input to the SIFT feature detection algorithm. \par

Colour optical microscopy (digital pathology) images (approximately 1500 x 1500 pixels) were converted to greyscale via the Python OpenCV package which implements the conversion function: \newline
\centerline{Y = 0.299*R + 0.587*G + 0.114*B} \newline \newline
where R, G and B represent the red, green and blue channel intensity values. The greyscale digital pathology images were then downsampled to match the resolution of the corresponding 500 x 500 Raman images.

\begin{figure}[h]
\centering
  \includegraphics[width=8.3cm]{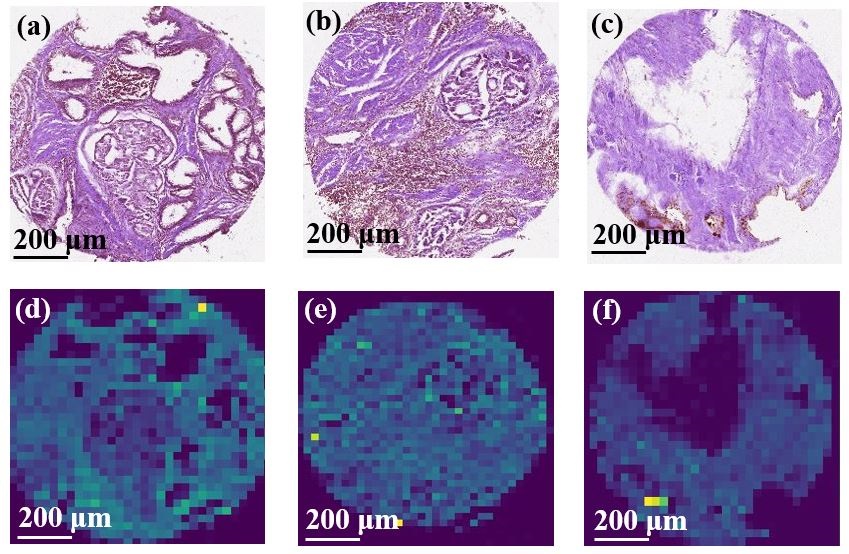}
  \caption{Two Gleason grade 4 samples ((a) and (b)) and a non-cancer sample (c). Colour digital histopathology images are shown in (a) to (c) with corresponding Raman images in (d) to (f).}
  \label{fgr:example}
\end{figure}

Visually, the upsampled greyscale Raman chemical images have low contrast values i.e. most of the pixel intensity values exist in a relatively narrow range per image. As a means of contrast stretching, histogram equalization was applied to each greyscale  Raman image. This transform flattens the greyscale image histogram in an effort to ensure that all intensity values are equally distributed. It is a common technique for normalizing image intensity prior to further processing. Figure 2 shows the image intensity histogram of a greyscale Raman chemical image before and after histogram equalization.

\begin{figure}[h]
\centering
  \includegraphics[width=8.3cm]{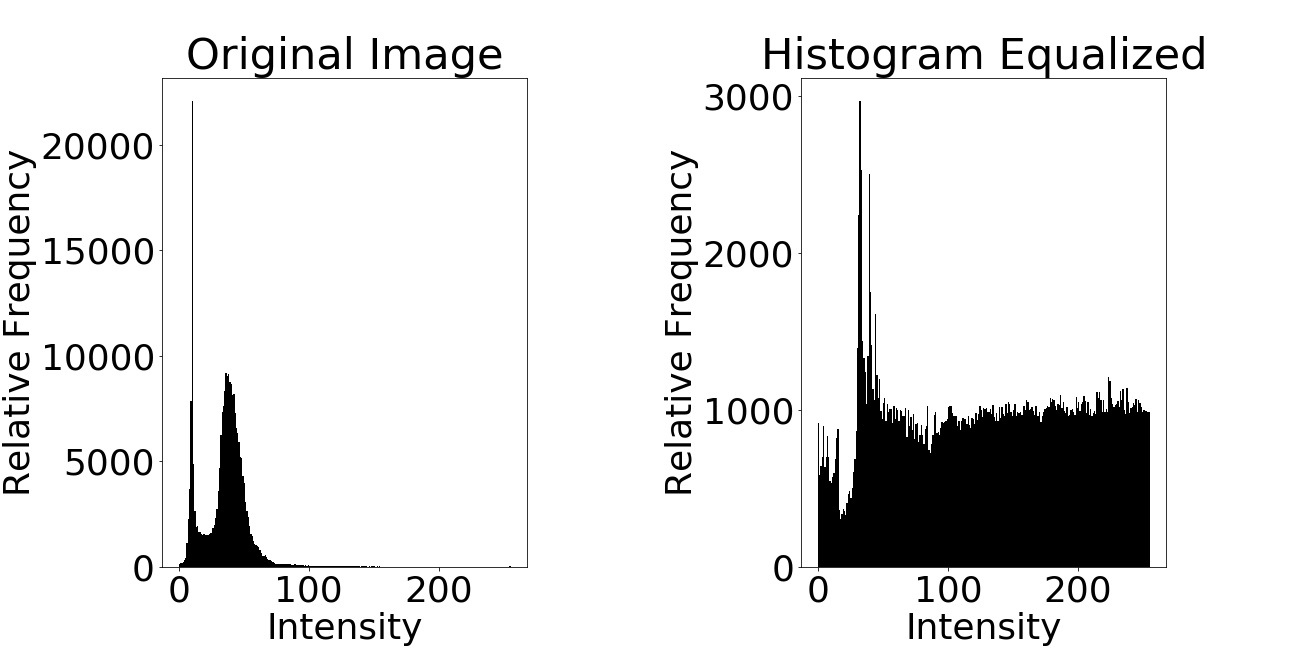}
  \caption{Image intensity histograms of greyscale Raman chemical image before (left) and after (right) application of histogram equalization.}
  \label{fgr:example}
\end{figure}

\subsection{Image-based model configurations}
Two binary classification tasks (non-cancer/cancer and Gleason grade 3/grade 4) were investigated for three imaging configurations. The first configuration was digital histopathology (optical microscopy), the second utilised RCI and the third was a combination of digital histopathology and RCI. Five-fold cross validation with each fold containing samples from disjoint sets of patients was investigated i.e patient sample independence was ensured. The feature extraction approach used in this study first partitions images into reference and classification sets. 
Reference images were used to form image features which then encode the classification set. 179 images were used for the binary non-cancer/cancer task - 52 reference images for defining the image features and 127 for building the classification models. 96 images were used for the Gleason grade 3/grade 4 task – 27 reference images and 69 images for the classification models. Both binary classification tasks were evaluated using a single reference and classification partition. On witnessing promising results for the task of Gleason grade 3/ grade 4 differentiation, modelling was extended to incorporate 10 different reference/classification partitions (average performance metrics over the 10 models were reported). This amounted to 9 model configurations in total. Table S3 in the Supplementary Material gives an overview of the models investigated.

\subsection{Models with Single \& Multiple Reference Sets}

SIFT feature encoding of the classification sets for models 1-6 in Table S3 were based on a single reference set. Reference sets were formed from samples belonging to a randomly selected set of patients. These patients and their samples were not used in the classification sets. The reference set of images were used to encode the remaining images (using the technique explained in the Feature Extraction section). These remaining images constituted the classification set and were used for model training and evaluation via cross-validation. Cross-validation was structured such that samples from 5 disjoint sets of patients were used in each fold - this ensured patient independence across folds.\par 
For models 7-9 in Table S3, 10 separate models based on differing reference/classification partitions were generated. Mean cross-validation performance scores were generated for each of the 10 models with the overall average of these individual means reported. As described above, the same approach to patient independence across the reference sets and cross-validation folds was followed.

\subsection{Reference - Classification Sets and Dictionary Sizes}

The aforementioned reference sets were used to form dictionaries of visual words for each model. Table 2 shows the breakdown of images used for both the reference and classification sets for each diagnostic classifier and pathological class (single reference set case).

\begin{table}[h]
\small
  \caption{\ Reference and Classification Set Image Distribution}
  \label{tbl:example}
  \begin{tabular*}{0.48\textwidth}{m{1.4cm} m{1.6cm} m{0.75cm} m{0.5cm} m{0.5cm} m{0.5cm} m{0.75cm}}
    \hline
    \textbf{Set} & \textbf{Diagnostic Model} & \textbf{Normal} & \textbf{G3} & \textbf{G4} & \textbf{G5} & \textbf{Total}\\
    \hline
    Reference & Non-cancer/ \newline Cancer & 23 & 16 & 12 & 1 &  52 \\
    & & & & & \\
     & Gleason 3/4 &  & 12 & 15 &  & 27\\
     \hline
    Classification & Non-cancer/ \newline Cancer & 59 & 35 & 33 & &  127 \\
    & & & & & \\
    & Gleason 3/4 &  & 39 & 30 & & 69\\
    \hline
  \end{tabular*}
\end{table}

The dictionary size parameter was optimized in a similar fashion to the optimisation of the support vector machine (SVM) hyperparameters. For the single imaging approach of digital histopathology, image histograms were generated and passed to the classification stage over a range of investigated dictionary sizes. These were [50, 75, 100, 200, 300, 500, 1000], which represent the number of visual words that were generated via the BoVW clustering stage and thus the visual words used to encode an image as a histogram of visual word frequencies. In the case of the RCI single imaging approach, the dictionary sizes used were [5, 10, 25, 50, 100, 200, 300].
For the multimodal approach, the range of dictionaries investigated were combinations of the values used for both single imaging approaches i.e. [50, 75, 100, 200, 300, 500, 1000] and [5, 10, 25, 50, 100, 200, 300] for the digital pathology and RCI images respectively. Image histograms of visual words were generated for each value in the tested ranges with permutations of these being used to optimise each model.

\subsection{Feature Extraction}

Scale Invariant Feature Transform (SIFT) was used to extract image features. SIFT is an image descriptor that can be used for image recognition. The descriptor shows invariance to scaling, rotation and translation in the image domain and exhibits robustness to moderate illumination variation and perspective transforms. The SIFT descriptor has been shown to be effective in practice for real-world object recognition. It utilizes a method that detects points of interest from grey-level images. Statistics relating to local gradient directions of image intensities are collected which give a description of local image structures in an interest point’s local neighbourhood.\cite{lindeberg2012scale} Figure 3 outlines the SIFT detection and description process.
The results of Scale Invariant Feature Transform yield high-dimensional feature vectors, each describing a single key point. Exhaustive matching of key point descriptors across images can be avoided by utilizing the classic bag-of-words paradigm from the field of information retrieval.\cite{csurka2004visual} The key point descriptors extracted by the SIFT algorithm are grouped into clusters with each cluster containing similar descriptors. Each cluster can be treated as a visual word that represents certain local patterns which are common among the descriptors in that cluster. The combination of these clusters yields a visual word vocabulary that describes a range of local image patterns. This mapping of key points to visual words represents a dictionary that can be used to represent the image as a bag of visual words.

\begin{figure}[h]
\centering
  \includegraphics[width=8.3cm,height=5.5cm]{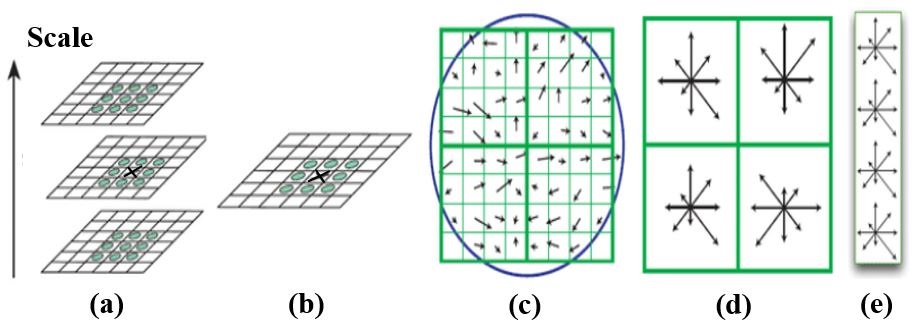}
  \caption{The SIFT detection and description process. (a) the difference of Gaussians is computed at multiple scales. X is a potential keypoint that is compared to its 8 neighbours at the same scale and 9 corresponding neighbouring pixels at each neighbouring scale ; (b) A scale is selected for each keypoint; (c) The Gaussian smoothed image at the keypoint’s scale is used to assign orientation. Gradient orientations are computed at that scale. The large circle represents the weighting of a Gaussian window when the orientation histograms are being calculated; (d) The orientations are spatially pooled; (e) This yields histograms that are concatenated and normalised to form the descriptor. Figure adopted from \cite{csurka2018handcrafted}.}
  \label{fgr:example}
\end{figure}

This equates to a vector containing the visual word count which can then be used at the classification stage.\cite{yang2007evaluating} In the case of the combined digital pathology and Raman imaging models, the feature vectors extracted from each modality were concatenated to form a single representation of the image pair for a tissue microarray sample. For the single modality cases, just the feature vector derived from the digital pathology or Raman chemical image was utilized. Figure 4 shows the main stages of the Bag of Visual Words approach.

\begin{figure*}[h]
\centering
  \includegraphics[width=17.1cm]{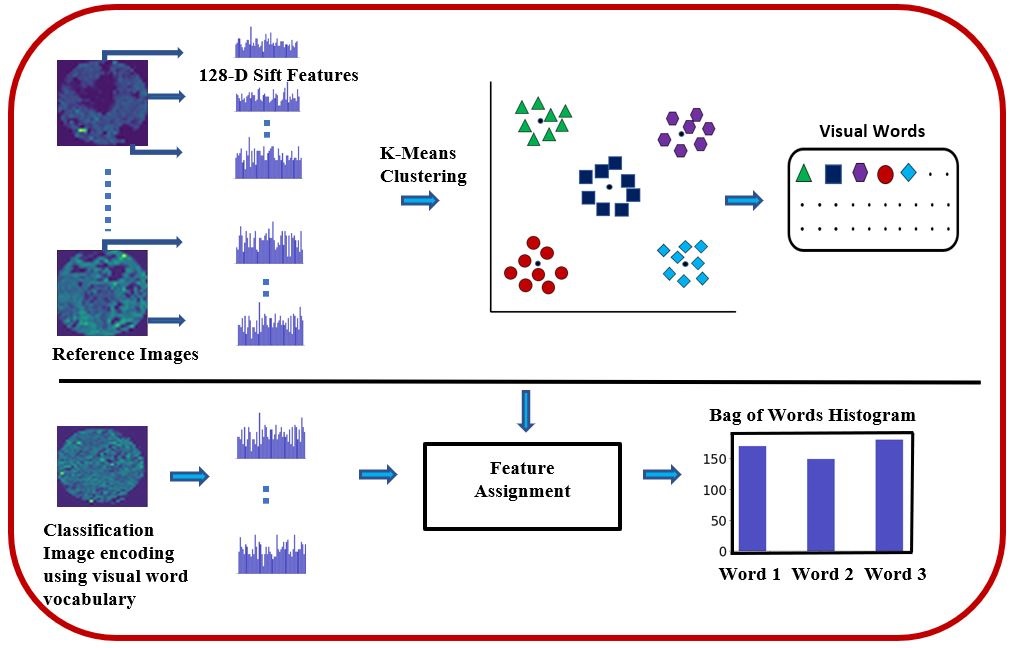}
  \caption{A diagram illustrating how to generate the bag-of-words visual words from several Raman Chemical Imaging reference images and how to represent an image by a bag-of-words histogram associated with these visual words. Figure adapted from \cite{lin2019two}.}
  \label{fgr:example}
\end{figure*}

The main steps outlined in Figure 4 involve: \newline
(a) The reference images are decomposed into many SIFT features. By K-Means clustering, these SIFT features are clustered to a predefined number of clusters. The centroids of these clusters are referred to as visual words. In this example, three visual words are generated. The set containing all the visual words is referred to as a dictionary e.g. 27 reference images were used to build a visual word dictionary for the Gleason grade 3/ grade 4 classification model.\newline
(b) Given a new image to be represented by a bag-of-words model, it is first decomposed into several SIFT features. Each SIFT feature is assigned to the nearest visual word using Euclidean distance i.e. vector quantization. Implementing such feature assignment for each SIFT feature, the image can be represented as a histogram of visual word frequencies i.e. the counts of the occurrences of the SIFT features assigned to each visual word.\cite{lin2019two} For the Gleason grade 3/4 classifier, 69 images were converted to visual word histograms based on the dictionary formed by the 27 reference images. These 69 images were then used to build a cross-validated classification model. The same process was applied for histopathology images.

\subsection{Image Histogram Visualisation}
The process of SIFT feature extraction and BoVW clustering yields image representations in the form of histograms of visual words. These frequency distributions detail how many of each type of feature (visual word) occurs in a given image. Fig. 5 shows examples of normal and cancer tissue sample BoVW image histograms for H\&E and Raman chemical imaging. Similarly, for the cases of Gleason grade 3 and 4, Fig. 6 contains examples of tissue sample BoVW image histograms pertaining to both image modalities.

\begin{figure}[h]
\centering
  \includegraphics[width=8.3cm]{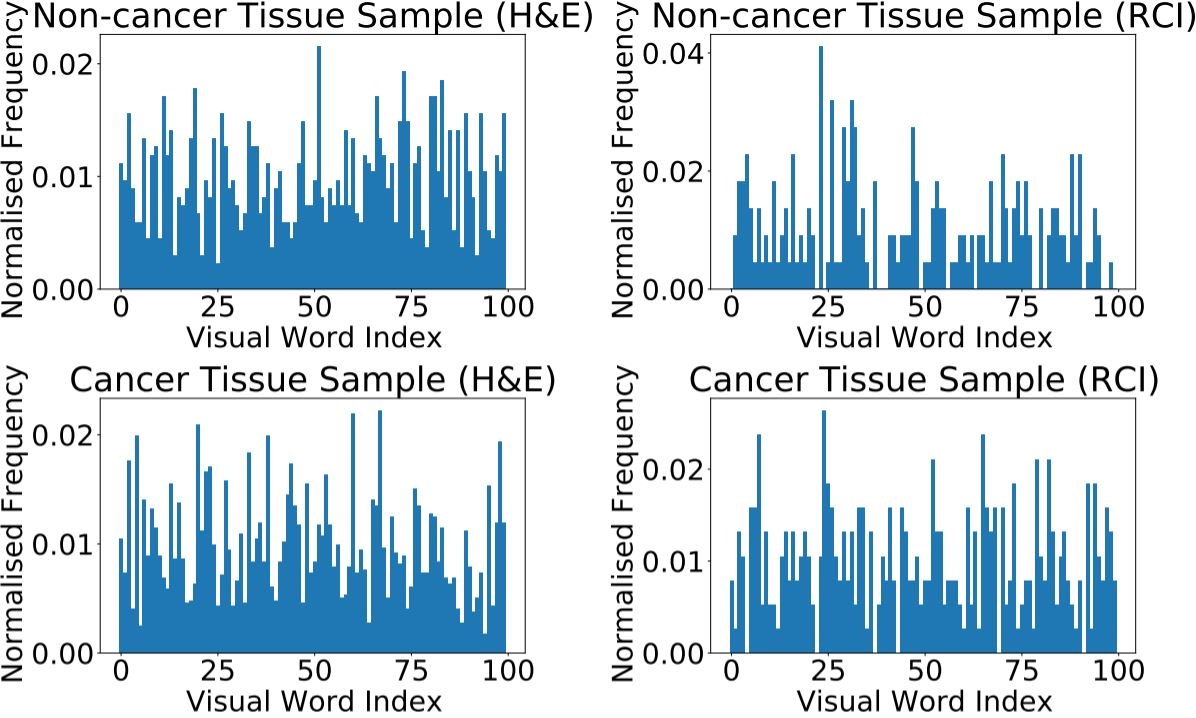}
  \caption{BoVW image histograms for a normal tissue sample H\&E image (top left)
with corresponding RCI image (top right) and a cancer tissue sample H\&E image 
(bottom left) with corresponding RCI image (bottom right).
}
  \label{fgr:example}
\end{figure}

\begin{figure}[h]
\centering
  \includegraphics[width=8.3cm]{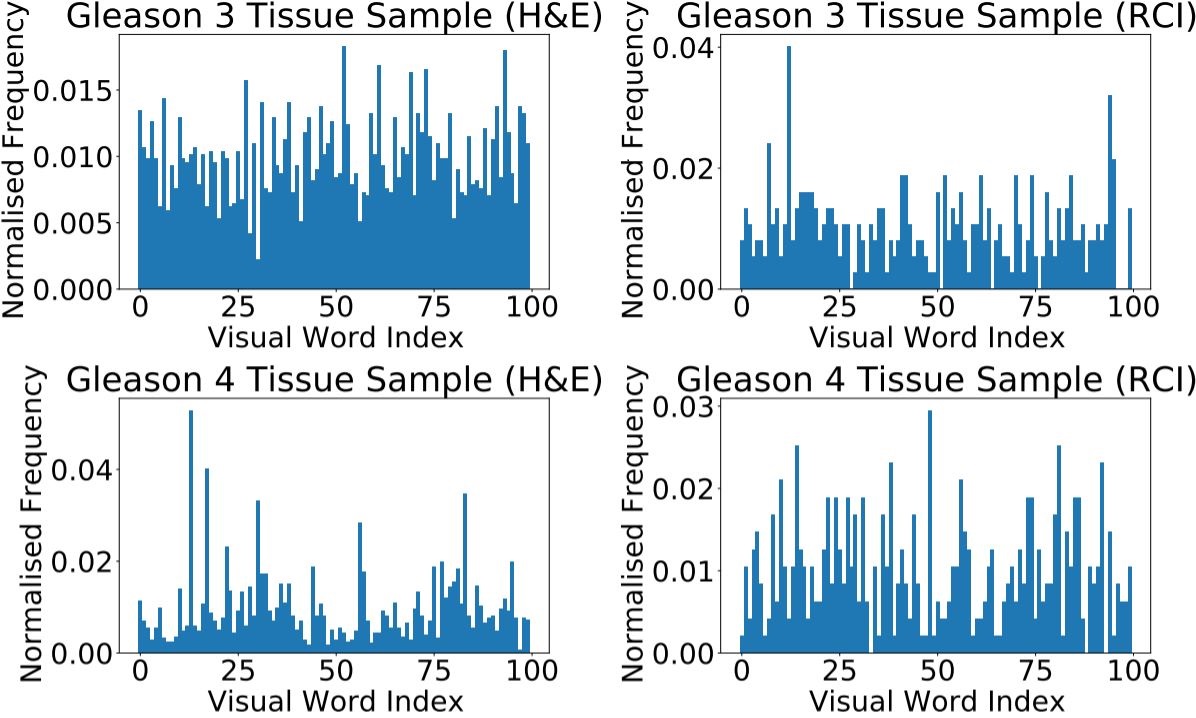}
  \caption{BoVW image histograms for a Gleason grade 3 tissue sample H\&E image   (top left) with corresponding RCI image (top right) and a Gleason grade 4 tissue sample H\&E image (bottom left) with corresponding RCI image (bottom right).}
  \label{fgr:example}
\end{figure}

\subsection{Image Feature Representations for Classification}

The histogram image representations generated in the feature extraction step are then passed to a classification algorithm. In the case of the combined digital pathology/RCI models, the feature vectors extracted from each modality were concatenated to form a single representation of the image pair for a tissue microarray sample. With the images converted to vector representations (histograms), a support vector machine was utilized to learn from training images and classify test images. Based on these image representations, the classifier outputs probability-like values that a tissue sample belongs to a given class.\par

Five-fold cross validation was used to assess the efficacy of models. This is a robust approach which partitions images into 5 approximately equal sized folds. Given that there were multiple sample images per patient, the folds of the cross-validation were formed from 5 disjoint sets of patient sample images. This approach ensures the independence of the training and test data sets within the cross-validation process. The overall algorithmic process is outlined in Figure 7.

\begin{figure*}[tbp]
\centering
  \includegraphics[width=17.1cm]{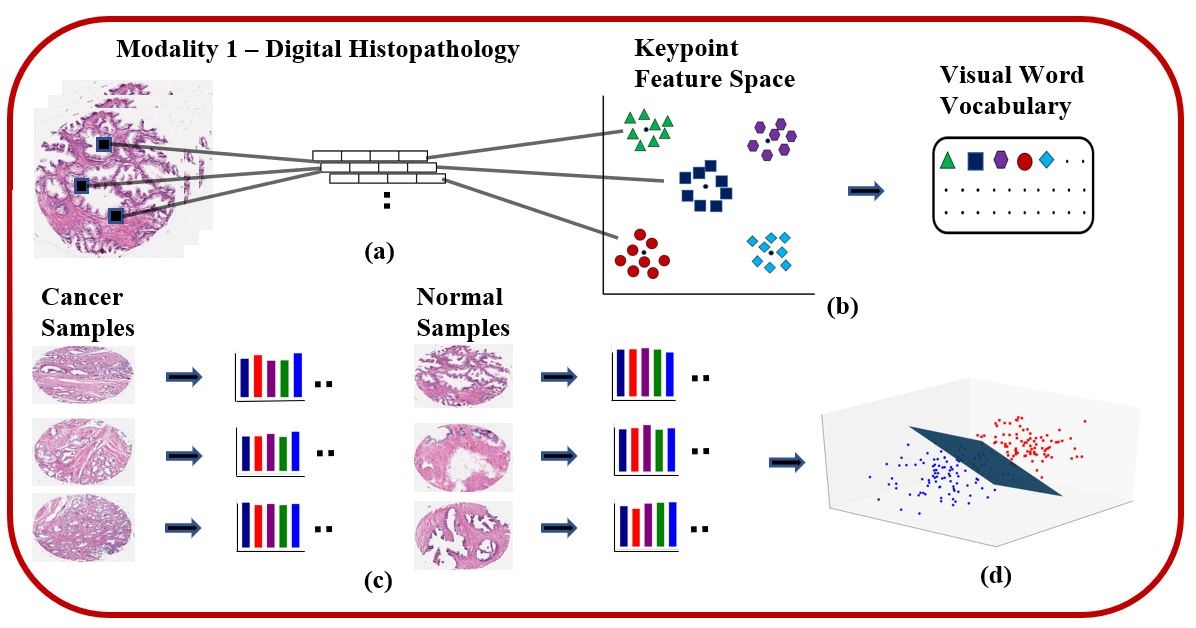}
  \label{fgr:example}
\end{figure*}

\begin{figure*}[tbp]
\centering
  \includegraphics[width=17.1cm]{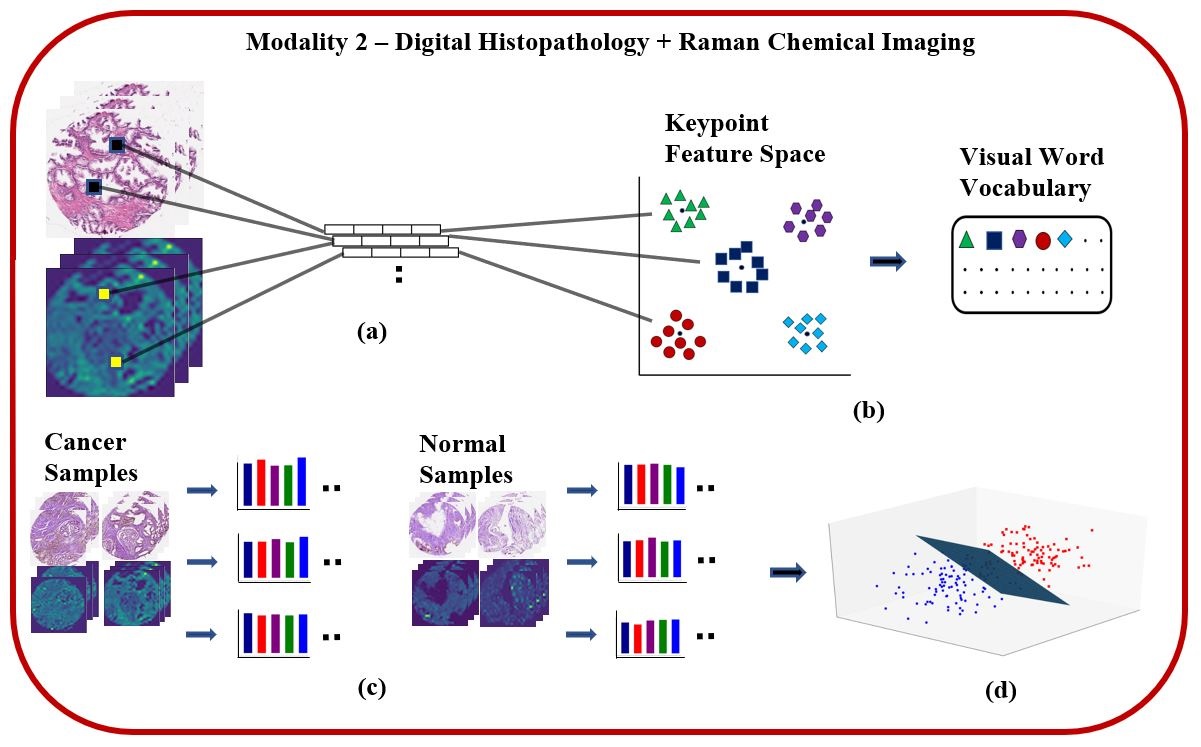}
\caption{Algorithmic process for feature extraction and classification for single and multimodal imaging cases. (a) Feature extraction (SIFT descriptors), (b) K-Means clustering yields visual words, (c) Feature quantization to create visual word histograms per image, (d) Classification stage. Figure adapted from \cite{suh2018sugar}.}
  \label{fgr:example}
\end{figure*}

\subsection{Classification with Support Vector Machines}

For the application of the SVM, the software package sklearn.svm from Python was used (version sklearn==0.0).\cite{pedregosa2011scikit} This involved implementation of the SVC (support vector classification) class which is based on libsvm. Two kinds of kernel functions were compared i.e. linear and radial basis function (RBF) in each experiment. Parameters were optimised by searching the four/five-dimensional grid which is created by ranges of C, $\gamma$, the chosen kernel and the SIFT/BoVW dictionary sizes for each imaging modality.

The permutation of these parameters which returns the maximum accuracy determines the optimal model. C is a penalty parameter for misclassification while $\gamma$ is related to the radius of influence of the support vectors that the model selects. The grid search process for parameter optimisation involved several steps:

1. Parameter C was sampled over the range [2\textsuperscript{-5}, 2\textsuperscript{-4},...,2\textsuperscript{15}] while {$\gamma$} was sampled over the range [2\textsuperscript{-15}, 2\textsuperscript{-14},...,2\textsuperscript{3}].\cite{hsu2003practical} The kernels tested were linear and radial basis function. Other model parameters used the default settings of the package.\par
2. The parameters given by the dictionary sizes for digital pathology and RCI were sampled over [50, 75, 100, 200, 300, 500, 1000] and [5, 10, 25, 50, 100, 200, 300] respectively. This equated to images being represented by histograms of varying length, equivalent to the number of visual words created via the clustering process of the BoVW dictionary generation.\par
3. Given the five parameters used for optimisation in the multimodal imaging approach i.e. digital pathology and RCI, the number of permutations searched is given by the product of the length of each parameter vector. In the case of the single modality model i.e. digital pathology imaging or RCI only, total permutations are given by the product of the length of four parameter value vectors i.e. C, $\gamma$, kernel choice and dictionary size. The process of parameter optimisation is discontinuous and attempts to find the global minimum for model error and maximum for model accuracy. It is unlikely that a true global minimum/maximum could be found given that a continuous search over unbounded ranges of the parameters is practically impossible. Additionally, a continuous differentiable function whose derivative exists at each point in the domain and that can be analytically solved is unknown.\par
4. Each parameter combination was used to derive image histograms of specified size and set SVM parameter values. Five-fold cross validation was used to assess the performance of the model with the metrics sensitivity, specificity, overall accuracy, AUC and the ROC curve being used to interpret results. Accuracy was optimised with the classes for each model being well balanced. Metrics from the cross-validation process were calculated as the mean values over each of the five folds. The probability-like values and categorical predictions obtained from the SVM classifier over each fold were used to form an ROC curve and confusion matrix for the entire dataset. Tables S4 and S5 outline the parameters of the optimised SIFT/BoVW SVM models.

\subsection{Classification using Raman spectral data}

For comparison with the image-based approaches, classification models were also constructed using median Raman spectra per sample, as described below.

\subsection{Calculation of median Raman spectrum}

For each Raman image collected, several pre-processing steps were carried out prior to calculation of the median Raman spectrum for each sample. First, in order to remove the image background (i.e. non-sample parts of the image), principal components analysis was applied to each image individually. The principal component (PC) score images were visually assessed, and the score image that best separated the sample from the background was selected. A mask was then created by thresholding that PC score image based on the distribution of pixel intensities in the PC score image histogram. Pixels containing cosmic rays and saturated spectral regions were subsequently removed using the approach described by Dorrepaal et al.\cite{dorrepaal2016tutorial}\par
Briefly, for each spectrum, standard normal variate pretreatment was applied to the spectrum; the difference in intensity between neighboring Raman shifts was calculated and the standard deviation of this ‘difference’ spectrum was calculated. Spectra with a standard deviation greater than the mean +/- 3 times the standard deviation of this value were identified as Cosmic Rays or saturated pixels and removed from the image by masking. Finally, in some image masks, very small regions outside of the core were noticed (possibly due to fragmentation of a core or residue from processing). In order to remove these from the images, all image regions with less than 10 pixels were identified and set to zero in the mask. A composite mask ( = background removal + cosmic ray/saturated pixels + small regions masks) was created and applied to the image. Subsequently the median Raman spectrum was calculated for each image.

\subsection{PLS-DA modelling}
Partial least squares discriminant analysis was applied to the median Raman spectra of the unstained TMAs for comparison with the multi-modal imaging approach. Models were constructed using untreated Raman spectra and six spectral pretreatments: SNV, 1st derivative Savitzky Golay pretreatment (‘1st der’,  window size = 15 points, polynomial order = 3), 2nd derivative Savitzky Golay pretreatment ((‘2nd der’,  window size = 15 points, polynomial order = 3), combinations of SNV followed by 1st or 2nd derivative pretreatment and multiplicative scatter correction (MSC). In addition, the fluorescence background was removed from the spectra using an exaggerated smoothing (0 order Savitzky Golay smoothing with a window size of 301) and subtracting the smoothed spectrum from the original median Raman spectrum. This pretreatment was then followed by the 6 spectral pretreatments described above.
Model parameters such as optimal pre-treatment and number of latent variables were selected by random cross validation on the data from the SIFT sets for non-cancer/cancer discrimination and Gleason grade 3/grade 4 discrimination. For random cross validation, 70\% of the spectra were randomly selected for model building and the remaining 30\% were used for cross validation (the ‘randperm’ function in MATLAB was used to randomly permute the data). This process was repeated 200 times and the mean accuracy, sensitivity and specificity were calculated for each combination of spectral pretreatment and number of latent variables. The optimal spectral pretreatment and number of latent variables was selected based on consideration of the mean accuracy, and product of sensitivity and specificity. Subsequently, 5-fold cross validation using the selected parameters was carried out using the same folds as used in the image based models. As described in Section 2.10, the cross-validation was structured such that samples from 5 disjoint sets of patients were used in each fold - this ensured patient independence across folds. Using the selected model parameters, calibration models were then re-built on the combination of the SIFT reference sets and the calibration set for a given fold number, and validated on the corresponding independent patient set. 
Unless otherwise specified, Raman image processing was carried out using MATLAB (MATLAB R2019a, The MathWorks Ltd) functions written in house and from the image processing toolbox.

\section{Results}
This study compared the performance of single modality approaches (digital histopathology, Raman chemical imaging (RCI) and median Raman spectra) to a multimodal (RCI + digital pathology) approach in classifying the pathological state of prostate tissue samples using SIFT and a support vector machine (SVM) classifier. A random set of patient's samples were used as the reference image set. Features from this set were clustered to yield visual words, which were then used to encode images in the classification set. Two binary classification tasks were explored in order to show the competing performance of the single and multimodal imaging approaches. These tasks were cancer/non-cancer and Gleason grade 3/grade 4 binary classification. Each of these tasks were tested with regard to both the single and multimodal imaging input types. The classification algorithm makes a diagnostic prediction based on the extracted features for each TMA core sample.

\subsection{Classification using Raman spectra}

Classification results for cancer vs. non-cancer based on PLS-DA models constructed on median Raman spectra are presented in the Supplementary material. The optimal PLS-DA model parameters (based on maximum product of sensitivity and specificity) were selected as no pretreatment followed by 2nd derivative Savitzky Golay pre-processing and a PLS-DA model with 9 latent variables. PLS-DA models using these parameters were then built and cross validated on the 5 folds used in the imaging models. \par
The resultant confusion matrix is shown in Figure 8. The accuracy of this approach was 58.3\%, with a sensitivity of 0.62 and a specificity of 0.54.

\begin{figure}[H]
\centering
  \includegraphics[width=8.1cm]{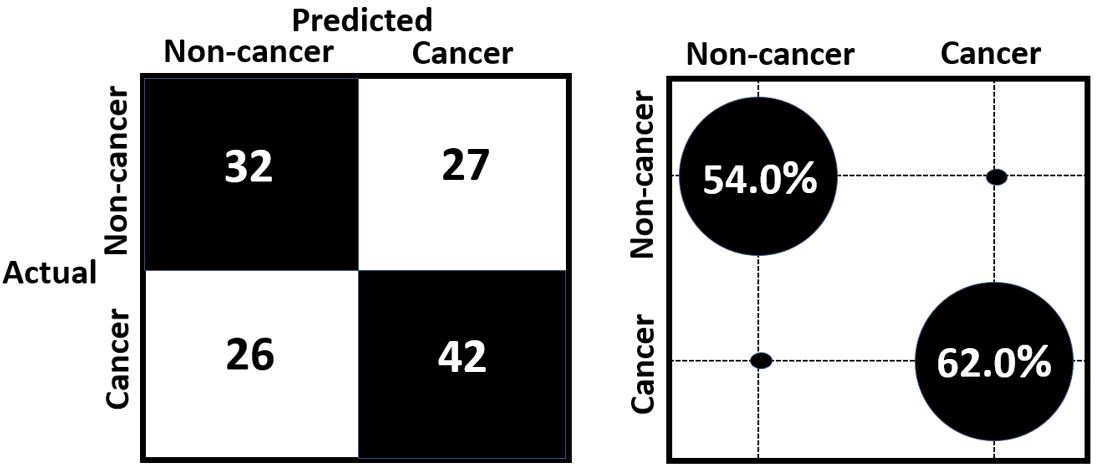}
  \caption{Confusion matrix for classification of cancer vs. non-cancer samples using Raman spectra (left) with confusion ball graphic of specificity and sensitivity (right). Confusion matrix based on pooled predictions over 5-fold cross-validation. Sensitivity and specificity are mean values over cross-validation folds.}
  \label{fgr:example}
\end{figure}

As for the PLS-DA models for Gleason grade 3/grade 4 classification, the product of mean sensitivity and specificity calculated over 200 random splits of the reference image set (shown in Figure S2) indicated optimal PLS-DA model parameters to be fluorescence removal followed by 1st derivative preprocessing and a PLS-DA model with 4 latent variables. PLS-DA models using these parameters were then built and cross validated on the same 5 folds used for the imaging models. The resultant confusion matrix is shown in Table S2. The accuracy of this approach was 42\%, with a sensitivity of 0.33 and a specificity of 0.49.\par 
These results show that the use of median Raman spectra alone produces relatively poor classification models, suggesting the need to consider the image related features in subsequent modelling. Therefore, in addition to a representative median spectra approach, we investigated a method based on the greyscale Raman chemical image of the tissue sample, calculated using the mean spectral value per pixel. This approach allowed the incorporation of both biochemical data from pixel spectra and structural and morphological information via the spatial distribution of pixel intensities.

\subsection{Classification using Raman Chemical Imaging}

Using RCI images, a 5-fold cross-validated model was constructed based on images encoded by a random reference set as detailed in Section 2.8 and Table 2. The binary non-cancer/cancer classifier yielded a sensitivity and specificity of 72.5\% $\pm$ 0.1\% and 69.0\% $\pm$ 0.1\% respectively (Figure 9) with a reported accuracy of 70.3\% $\pm$ 2.8\% and AUC of 0.707 $\pm$ 0.038.

\begin{figure}[H]
\centering
  \includegraphics[width=8.1cm]{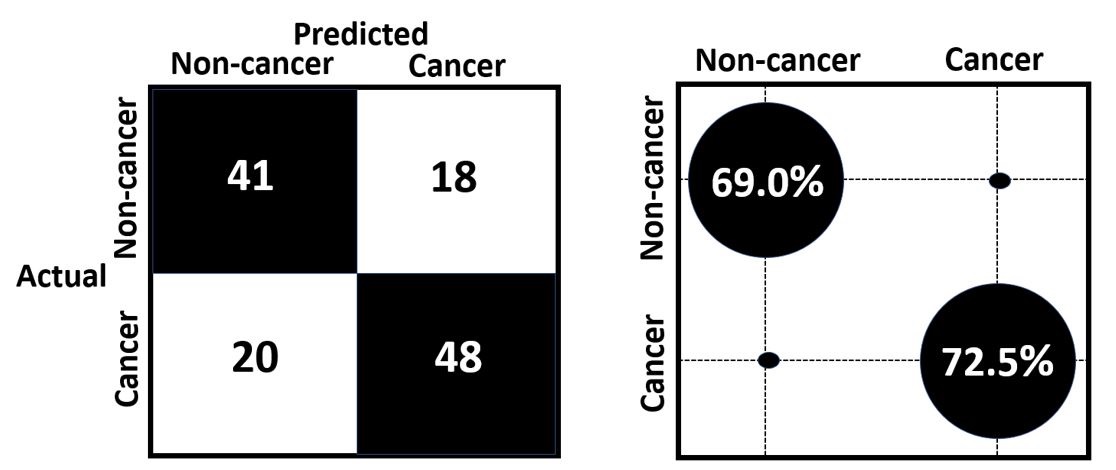}
  \caption{Confusion matrix for classification of cancer vs. non-cancer samples using
RCI imaging alone (left) with confusion ball graphic of specificity and sensitivity (right). Confusion matrix based on pooled predictions over 5-fold cross-validation. Sensitivity and specificity are mean values over cross-validation folds.
}
  \label{fgr:example}
\end{figure}

Gleason grade 3 and 4 classification was then investigated using RCI imaging. Similarly, a 5-fold cross-validated model with images encoded by a random reference set was investigated. The model achieved an accuracy of 68.0\% $\pm$ 20.0\%, sensitivity of 72.7\% $\pm$ 22.5\% and specificity of 68.0\% $\pm$ 29.5\% (Figure 10). The AUC was found to be 0.750 $\pm$ 0.192. Results in this section were mean values per fold.

\begin{figure}[H]
\centering
  \includegraphics[width=8.1cm]{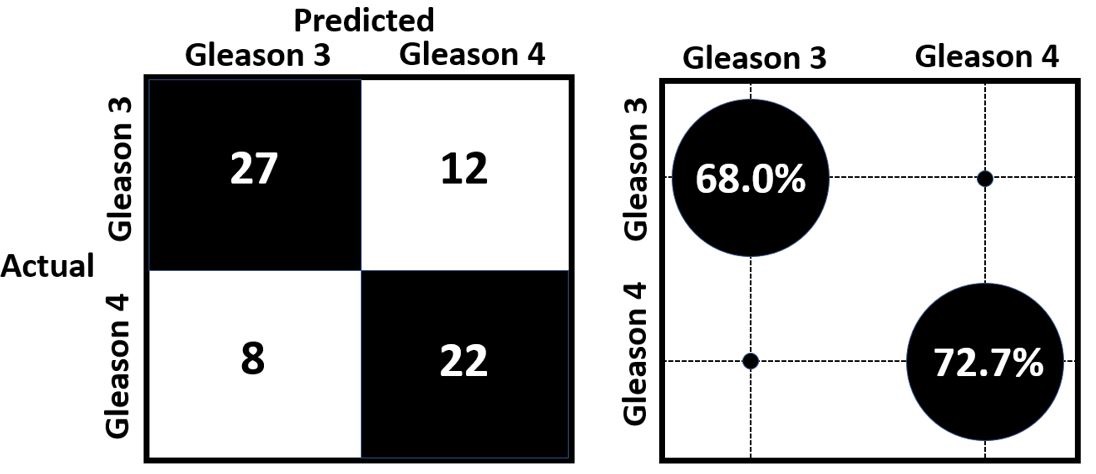}
  \caption{Confusion matrix for classification of Gleason grades 3 and 4 samples
using RCI imaging alone (left) with confusion ball graphic of specificity and sensitivity (right). Confusion matrix based on pooled predictions over 5-fold cross-validation. Sensitivity and specificity are mean values over cross-validation folds.
}
  \label{fgr:example}
\end{figure}

\subsection{Classification using Digital Pathology Imaging}

Using digital pathology images, a 5-fold cross-validated model was constructed based on images encoded by a random reference set. The binary classification of cancer and non-cancer samples yielded a sensitivity and specificity of 81.9\% $\pm$ 15.3\%  and 80.0\% $\pm$ 16.4\% respectively (Figure 11) with a reported accuracy of 82.3\% $\pm$ 6.0\% and AUC of 0.878 $\pm$ 0.058.

\begin{figure}[H]
\centering
  \includegraphics[width=8.1cm]{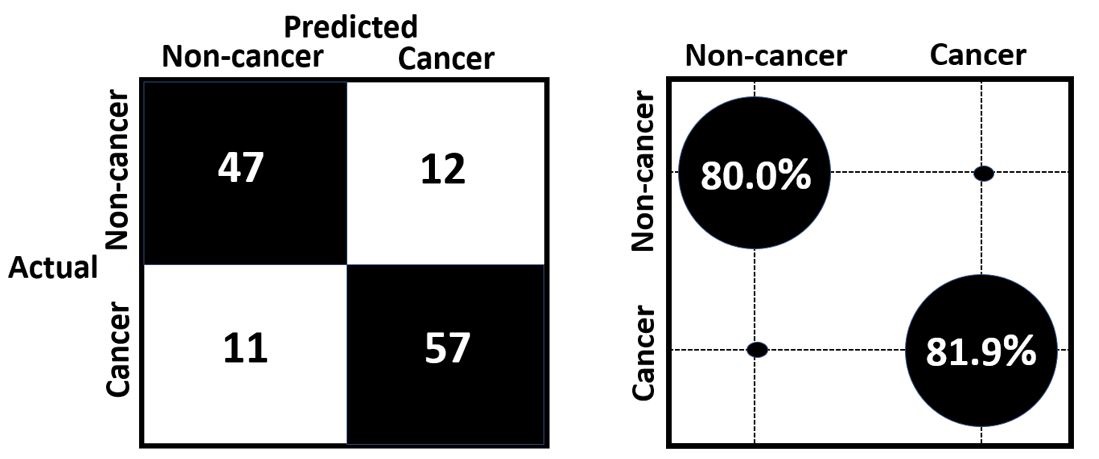}
  \caption{Confusion matrix for classification of cancer vs. non-cancer samples using digital pathology imaging alone (left) with confusion ball graphic of specificity and sensitivity (right). Confusion matrix based on pooled predictions over 5-fold cross-validation. Sensitivity and specificity are mean values over cross-validation folds.
}
  \label{fgr:example}
\end{figure}

The binary classification of Gleason grade 3 and 4 was investigated using only features from digital pathology images. Applying a 5-fold cross-validated model with images encoded by a random reference set, an accuracy of 74.1\% $\pm$ 9.3\%, sensitivity of 54.1\% $\pm$ 30.0\% and specificity of 84.7\% $\pm$ 3.0\% was achieved (Figure 12). AUC was given as 0.731 $\pm$ 0.211.
Results in this section were mean values per fold.

\begin{figure}[H]
\centering
  \includegraphics[width=8.1cm]{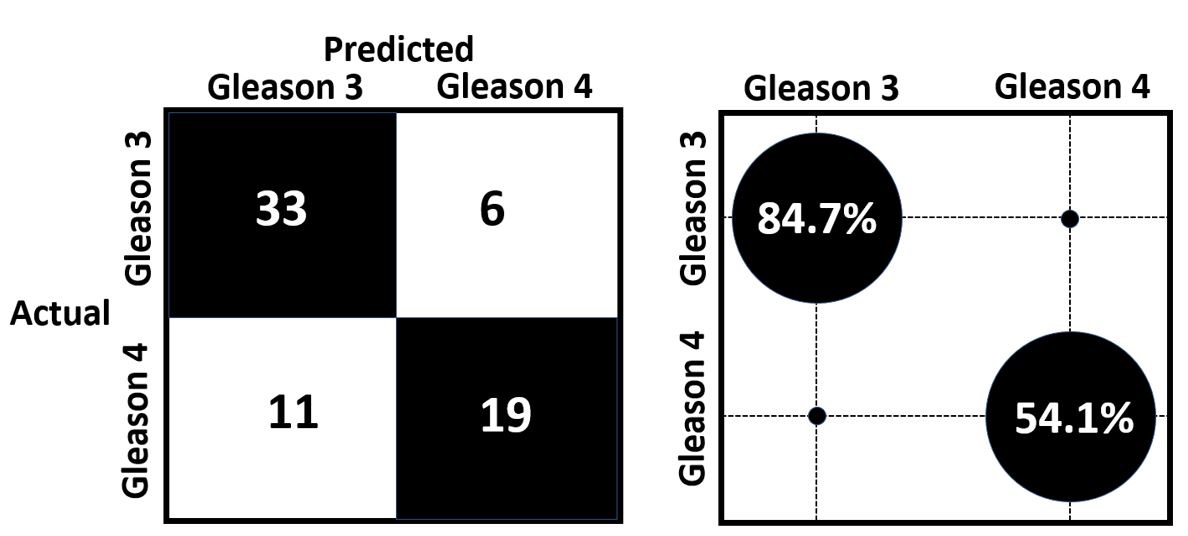}
  \caption{Confusion matrix for classification of Gleason grades 3 and 4 samples using digital pathology imaging alone (left) with confusion ball graphic of specificity and sensitivity (right). Confusion matrix based on pooled predictions over 5-fold cross-validation. Sensitivity and specificity are mean values over cross-validation folds.
}
  \label{fgr:example}
\end{figure}

\subsection{Classification using Multimodal Imaging}

The multimodal model of digital pathology and Raman imaging applied to classification of non-cancer/cancer samples yielded a sensitivity and specificity of 78.4\% $\pm$ 18.7\% and 81.6\% $\pm$ 6.9\% respectively on 5-fold cross-validation with a random reference set for encoding classification images (Figure 13). Accuracy was found to be 80.8\% $\pm$ 7.7\% with AUC being 0.845 $\pm$ 0.065.

\begin{figure}[H]
\centering
  \includegraphics[width=8.1cm]{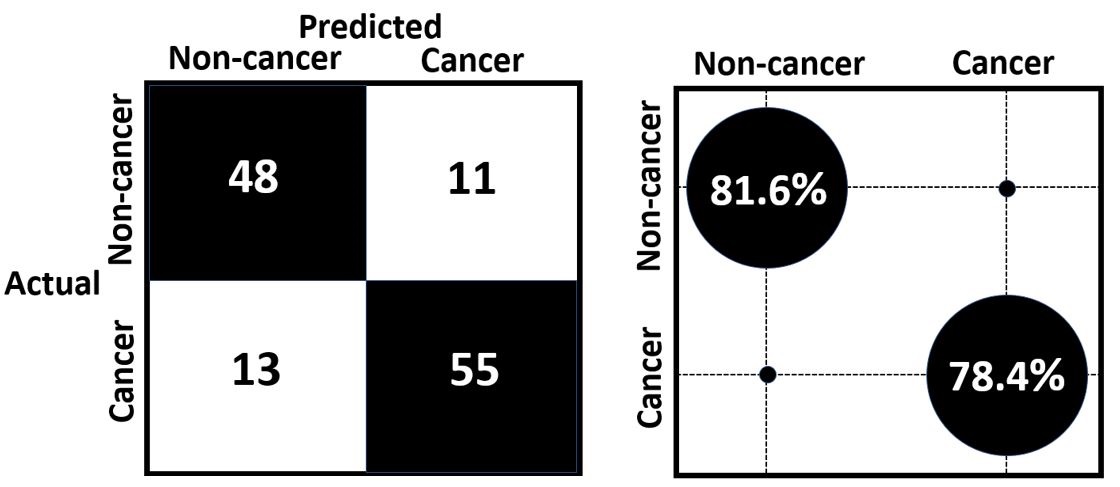}
  \caption{Confusion matrix for classification of cancer vs. non-cancer samples using combined digital pathology and Raman imaging (left) with confusion ball graphic of specificity and sensitivity (right). Confusion matrix based on pooled predictions over 5-fold cross-validation. Sensitivity \& specificity are mean values over cross-validation folds.
}
  \label{fgr:example}
\end{figure}

In the case of Gleason grade 3 and 4 classification, a model utilising features from both digital pathology and Raman imaging modalities gave an accuracy, sensitivity and specificity of 82.8\% $\pm$ 18.1\%, 73.8\% $\pm$ 27.3\% and 88.1\% $\pm$ 16.8\% respectively for 5-fold cross-validation, with images encoded by a random reference set (Figure 14). The AUC was 0.858 $\pm$ 0.186. Results in this section were mean values per fold.

\begin{figure}[H]
\centering
  \includegraphics[width=8.1cm]{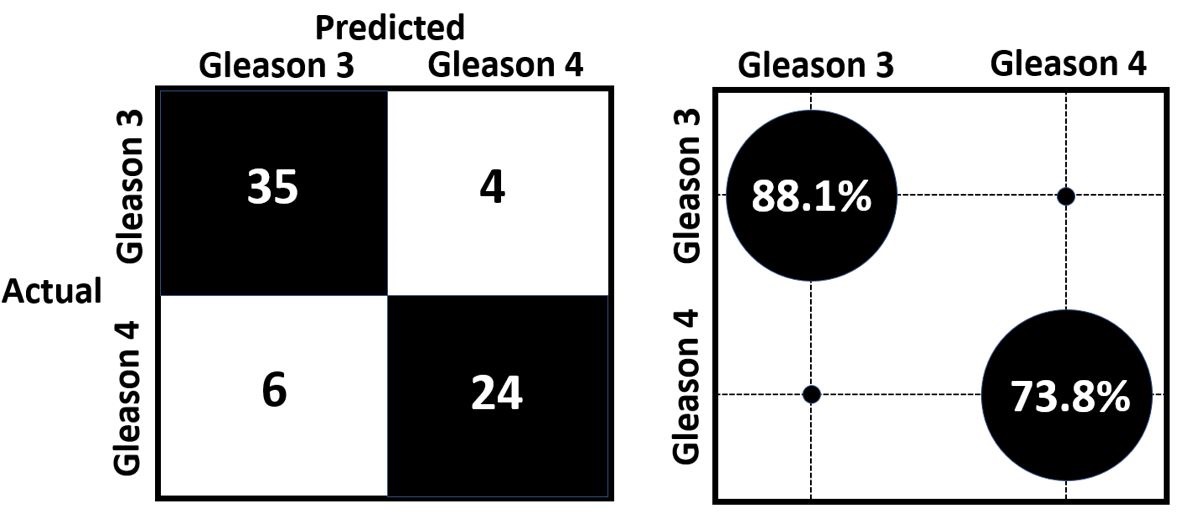}
  \caption{Confusion matrix for classification of Gleason grades 3 and 4 samples using combined digital pathology and Raman imaging (left) with confusion ball graphic of specificity and sensitivity (right). Confusion matrix based on pooled predictions over 5-fold cross-validation. Sensitivity \& specificity are mean values over cross-validation folds.
}
  \label{fgr:example}
\end{figure}

\subsection{ROC Curves Comparison}
For the binary classification of cancer and non-cancer samples, the ROC curves for Raman chemical imaging, digital pathology and the multimodal combination of both imaging modalities can be seen in Figure 15. The receiver operator characteristic (ROC) curves indicate that models based on the single modality of digital pathology and multimodal combination of digital pathology and Raman chemical imaging exhibit closely matched diagnostic ability. \par
From observation of the ROC curve in Figure 15, it can be seen that the digital pathology model (red curve) generally achieves slightly better sensitivity than the multimodal model with the exception of the region where specificity ranges between 80\% and 90\%. RCI does exhibit some ability to differentiate sample classes but appears to add little when combined with the digital pathology approach for the non-cancer/cancer task.

\begin{figure}[H]
\centering
  \includegraphics[width=8.1cm]{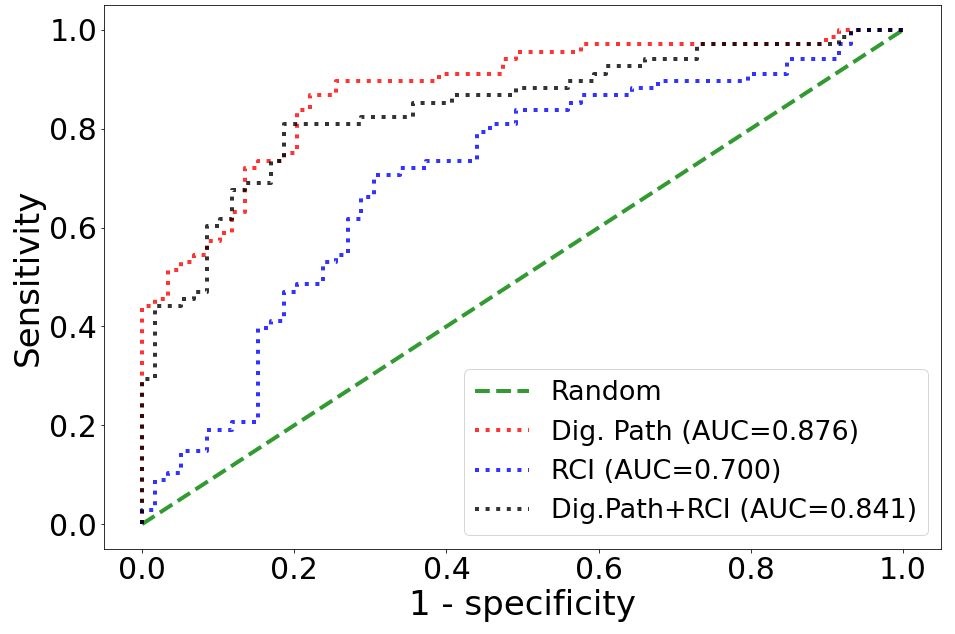}
  \caption{Receiver operator characteristic (ROC) comparison of binary cancer/non-cancer classifiers for digital pathology, Raman chemical imaging and the multimodal combination of both imaging types. Curves and AUC values are based on pooled model predictions from 5-fold cross-validation.
}
  \label{fgr:example}
\end{figure}

For the classification of Gleason grade 3 and 4 samples, ROC curves for digital pathology, Raman chemical imaging and the multimodal combination of both are shown in Figure 16. The improvement in diagnostic ability when augmenting digital pathology with RCI is evident. The AUC values in both ROC graphs are generated from the pooled predictions over the 5 cross-validation folds which accounts for the difference when compared with the mean AUC values per fold. 

\begin{figure}[H]
\centering
  \includegraphics[width=8.1cm]{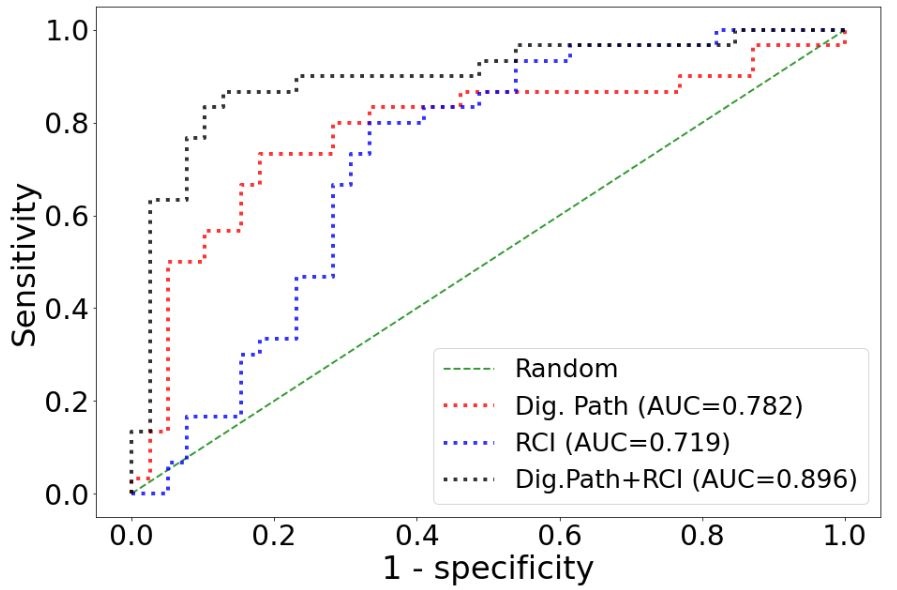}
  \caption{Receiver operator characteristic (ROC) comparison of Gleason grade 3 and 4 classifiers for digital pathology, Raman chemical imaging and the multimodal combination of both imaging types.Curves and AUC values are based on pooled model predictions from 5-fold cross-validation.
}
  \label{fgr:example}
\end{figure}

\subsection{Multiple Reference Set Results and T-test}

Paired t-tests were conducted to ascertain if there was a significant difference in the AUC values of the digital pathology and multimodal models presented in the preceding sections. The AUC values were those calculated on the validation sets of the 5-fold cross-validation of both models. No significant difference was observed between the digital pathology and multimodal approaches for the binary non-cancer/cancer classification task. The mean AUC difference (M=0.033, SD=0.064, N=5) was not significantly greater than zero, t(5)=2.776, two tail p=0.318. However, in the case of Gleason grade 3/grade 4 classification, the fusion of digital pathology and RCI showed potential for improved diagnostic performance for prostate cancer severity. The mean AUC difference (M=0.127, SD=0.059, N=5) was significantly greater than zero, t(5)=2.776, two tail p=0.009, providing evidence that the model is effective in improving binary classification of Gleason grade 3 and 4 prostate cancer. The 95\% confidence interval about the mean difference is (0.054, 0.201).

Results pertaining to the Gleason grade 3/grade 4 classification task where the average performances were reported over ten models (each model with different random patient groupings for the reference set and cross-validation folds) are contained in Table 3.

\begin{table}[h]
\small
  \caption{\ Average results of 5-fold cross-validation over 10 models (randomly selected patient groups for reference set and cross-validation folds) for Gleason grade 3/4 classification across digital pathology (DP), Raman Chemical Imaging (RCI) and multimodal approaches (DP + RCI)}
  \label{tbl:example}
  \begin{tabular*}{0.48\textwidth}{m{2cm} m{1.1cm} m{1.1cm} m{1.1cm} m{1.1cm}}
    \hline
    \textbf{Imaging Modality} & \textbf{Sensitivity} & \textbf{Specificity} & \textbf{AUC} & \textbf{False Negative Rate}\\
    \hline
    RCI & 63.3\% & 77.8\% & 0.714 & 36.7\% \\
    DP & 63.5\% & 84.9\% & 0.764 & 36.5\% \\
    Multimodal & 74.9\% & 83.4\% & 0.826 & 25.1\% \\
    \hline
  \end{tabular*}
\end{table}


It was noted that there was generally a reduced advantage between the multimodal and digital pathology model for Gleason grade 3 and 4 classification when taking the average results of 10 different configurations of reference set and cross-validation folds (Table 3), as compared to results using a single reference set in Sections 3.3 and 3.4. 

\section{Discussion}

Integration of Raman spectroscopy with other modalities can lead to significantly improved performance of a combined multimodal measurement in comparison to what either modality could achieve alone.\cite{das2017raman} Raman spectroscopy and its derivations are valuable for a range of clinical applications in the field of medicine including prostate cancer.\cite{kast2014emerging} To date, few studies have investigated the value in combining Raman Chemical Imaging (RCI) with other modalities. This work explores the integration of digital histopathology with RCI for the classification of prostate cancer tissue samples. We investigated whether a machine learning algorithm could perform better using this multimodal approach in comparison to utilising a range of single imaging inputs. Sample labels defined from pathologist examination of H\&E images were applied to the corresponding RCI images.\par
The developed machine learning algorithm predicts the pathological grade of the sample based on extracted image features. Two binary classification models were developed: the first classifies tissue samples as cancer or non-cancer with the second differentiating malignant samples as Gleason grade 3 or 4. The study compared performances of models utilizing the single modalities of digital histopathology, RCI and median Raman spectra in addition to the multimodal combination of  digital histopathology and RCI. It was observed that the multimodal imaging approach outperformed single imaging for the differentiation of Gleason grades 3 and 4. The best binary non-cancer/cancer models were those utilising digital pathology images and multimodal digital pathology/RCI images, with both achieving similar performance. Given that binary non-cancer/cancer diagnosis is a simpler task than differentiation of Gleason grade 3 and 4, there may have been enough information in the histopathological sample images to preclude any advantage from multimodal imaging. However, as pathologists can find it more challenging to differentiate between Gleason grade 3 and 4, the additional sub-visual information from the unstained RCI images may explain the improved diagnostic performance observed in this case. \par
Differentiation of Gleason grades 3 and 4 is a clinically relevant problem with a potentially high cost of error in assessment of these patterns. Where the discovery of a Gleason pattern 4 on follow-up biopsy would prompt the termination of active surveillance for a patient, misclassification of Gleason patterns 3 and 4 would have a major therapeutic impact.\cite{mckenney2011potential} In the case of Gleason grade 3 and 4 classification, a paired t-test indicated that the mean difference in area under the ROC curve for cross-validation fold values was significantly greater than zero where P<0.05 was considered statistically significant. This observation implies that there are properties and features of the RCI image that augment the information derived from histopathological images. This may be a reasonable assumption given that RCI produces a biochemical map of the sample with the intensities of the mean Raman image deriving directly from the spectral contributions of various biological components such as DNA, RNA and the extracellular matrix. Additionally, the spatial property of RCI contributes morphological information by accounting for changes in intensities within local neighbourhoods and thus identifying structural features within the sample.\par 
A further finding of note is the relatively good performance of the Gleason grade 3/4 classification model that used RCI imaging only as input. This suggests potential to use mean Raman images for Gleason grade 3/4 diagnosis. Less processing (such as chemical staining of samples) would be required with a few spectral points needed to form the chemical image. A lower cost/relatively rapid Raman-based device could potentially be constructed using this principle.\par
In this study, limited samples precluded the use of a separate test set in addition to the cross-validation approach used. The classification algorithms would benefit from acquisition of more data with the increased training likely leading to greater robustness and improved predictive power. The study is a preliminary validation; to improve chances of developing a clinically useful diagnostic classifier, the proposed multimodal imaging model should be trained on samples from a significant cohort of patients and evaluated on data from a range of hospitals and laboratories. This would introduce greater variation and better reflect real world data that the diagnostic classifier would encounter.\par

The primary stages of the model were image preprocessing, feature extraction and classification. Histogram equalization, SIFT with visual BoW and an SVM classifier were utilized to propose a model that could serve in diagnosis of prostate cancer pathological grades within a digital histopathological and Raman chemical imaging workflow. Support vector machines are commonly used due to their computational efficiency and robustness as they minimize over-fitting by adding structural constraints. Despite these properties and the use of stratified K-fold cross validation, additional validation on larger datasets and the use of separate test sets are required to further assess the generalizability of our models. Other limitations of the study include the use of multiple samples per patient. To ensure patient independence in the modelling process, samples from disjoint sets of patients were used for the reference and cross-validation folds. In our study, labelling is done by a single pathologist, which may carry greater misclassification risk versus having multiple labelers. The employment of a consensus pathology approach should strengthen the ground truth.\par
With acquisition of extended data, future work would include the application of deep learning and transfer learning to assess RCI augmentation of digital pathology and the use of methods that facilitate some degree of explainable artificial intelligence so that the model diagnostic decision process could be interrogated. The addition of a tumour segmentation step is worth exploring as many studies benefit from guiding the feature extraction process to specific areas of the image.

\section{Conclusion}

Multimodal image analysis is an important area of biomedical science. The combination of information from multiple sources can lead to improved diagnosis and treatment strategies, directly benefiting the patient’s quality of life and prognosis. This work demonstrates that the fusion of digital histopathology and Raman chemical imaging modalities has potential to improve the diagnostic performance for prostate cancer by integrating both morphological and biochemical information across both data sources. Comparative results across the single and multimodal approaches to non-cancer/cancer and Gleason grade 3/grade 4 differentiation indicates that a multimodal approach does not improve tumour identification but is useful for the more challenging task of tumour severity identification.\par 
The reasonable performance of approaches based on Raman chemical imaging only, indicates potential for a cost-effective and rapid means of assessing prostate cancer grades without requiring staining. However, while this study suggests that the integration of digital histopathology and Raman chemical imaging is a promising research method for the critical and challenging diagnosis of Gleason grade 3 and 4, the observed modest improvement should be considered in the context of the current cost and time required for Raman chemical imaging.

\section{Author Contributions}
AG, WMG, RWGW and SPF contributed to experimental planning
and design. EK was involved in proposal development and academic supervision. NAA designed and developed the experimental methodology for TMA Raman mapping acquisition and production of chemical images, described the experimental methodology, developed a protocol for sample preparation and curated data. AR contributed via protocol development and preparation for H\&E samples and data interpretation, while AON was involved in tissue sample and TMA preparation. TM acquired and processed Raman spectra and images. TD carried out image processing, conducted formal analysis through development of the machine learning methodology and classification models and was involved in overall manuscript conceptualisation and writing. CA annotated and labelled digital histopathology TMA core samples. SMK advised on machine learning and classification model
development and was involved in academic supervision. PJ contributed to solution engineering and design. AG conducted formal
analysis of Raman spectra models and contributed to manuscript writing. SPF, WMG, RWGW, SMK, PJ and AG reviewed and edited the final manuscript.

\section{Conflicts of interest}
WMG is co-founder, shareholder and Chief Scientific Officer of OncoMark Limited, as well as Scientific Advisory Board member for Carrick Therapeutics.

\section{Acknowledgements}

This research has been funded by the Irish Health Research Board (Grant Number HRA-POR-2015-1078), with additional support from the Science Foundation Ireland Investigator Programme OPTi-PREDICT (grant code 15/IA/3104); and the Science Foundation Ireland Strategic Partnership Programme Precision Oncology Ireland POI (grant code 18/SPP/3522). Funding for the Irish Prostate Cancer Research Consortium tissue samples was from Science Foundation Ireland, Grant number: TRA/2010/18; Irish Cancer Society, Grant number: PCI11WAT; Health Research Board, Grant number: TRA/2010/18; Welcome Trust-HRB Dublin Centre for Clinical Research.



\balance


\bibliography{rsc} 

\providecommand*{\mcitethebibliography}{\thebibliography}
\csname @ifundefined\endcsname{endmcitethebibliography}
{\let\endmcitethebibliography\endthebibliography}{}
\begin{mcitethebibliography}{79}
\providecommand*{\natexlab}[1]{#1}
\providecommand*{\mciteSetBstSublistMode}[1]{}
\providecommand*{\mciteSetBstMaxWidthForm}[2]{}
\providecommand*{\mciteBstWouldAddEndPuncttrue}
  {\def\EndOfBibitem{\unskip.}}
\providecommand*{\mciteBstWouldAddEndPunctfalse}
  {\let\EndOfBibitem\relax}
\providecommand*{\mciteSetBstMidEndSepPunct}[3]{}
\providecommand*{\mciteSetBstSublistLabelBeginEnd}[3]{}
\providecommand*{\EndOfBibitem}{}
\mciteSetBstSublistMode{f}
\mciteSetBstMaxWidthForm{subitem}
{(\emph{\alph{mcitesubitemcount}})}
\mciteSetBstSublistLabelBeginEnd{\mcitemaxwidthsubitemform\space}
{\relax}{\relax}

\bibitem[Stewart \emph{et~al.}(2016)Stewart, Wild, for Research~on
  Cancer,\emph{et~al.}]{stewart20162014}
B.~Stewart, C.~Wild, I.~A. for Research~on Cancer \emph{et~al.},
  \emph{WHO.(2014) World Cancer Report 2014}, 2016\relax
\mciteBstWouldAddEndPuncttrue
\mciteSetBstMidEndSepPunct{\mcitedefaultmidpunct}
{\mcitedefaultendpunct}{\mcitedefaultseppunct}\relax
\EndOfBibitem
\bibitem[Siegel \emph{et~al.}(2020)Siegel, Miller, and Jemal]{siegel2020cancer}
R.~L. Siegel, K.~D. Miller and A.~Jemal, \emph{CA: a cancer journal for
  clinicians}, 2020, \textbf{70}, 7--30\relax
\mciteBstWouldAddEndPuncttrue
\mciteSetBstMidEndSepPunct{\mcitedefaultmidpunct}
{\mcitedefaultendpunct}{\mcitedefaultseppunct}\relax
\EndOfBibitem
\bibitem[Humphrey
  \emph{et~al.}(2003)Humphrey\emph{et~al.}]{humphrey2003prostate}
P.~A. Humphrey \emph{et~al.}, \emph{Prostate pathology}, American Society for
  Clinical Pathology Chicago, 2003\relax
\mciteBstWouldAddEndPuncttrue
\mciteSetBstMidEndSepPunct{\mcitedefaultmidpunct}
{\mcitedefaultendpunct}{\mcitedefaultseppunct}\relax
\EndOfBibitem
\bibitem[Lomas and Ahmed(2020)]{lomas2020all}
D.~J. Lomas and H.~U. Ahmed, \emph{Nature Reviews Clinical Oncology}, 2020,
  1--10\relax
\mciteBstWouldAddEndPuncttrue
\mciteSetBstMidEndSepPunct{\mcitedefaultmidpunct}
{\mcitedefaultendpunct}{\mcitedefaultseppunct}\relax
\EndOfBibitem
\bibitem[Gleason and Mellinger(1974)]{gleason1974prediction}
D.~F. Gleason and G.~T. Mellinger, \emph{The Journal of urology}, 1974,
  \textbf{111}, 58--64\relax
\mciteBstWouldAddEndPuncttrue
\mciteSetBstMidEndSepPunct{\mcitedefaultmidpunct}
{\mcitedefaultendpunct}{\mcitedefaultseppunct}\relax
\EndOfBibitem
\bibitem[BenTaieb and Hamarneh(2017)]{bentaieb2017adversarial}
A.~BenTaieb and G.~Hamarneh, \emph{IEEE transactions on medical imaging}, 2017,
  \textbf{37}, 792--802\relax
\mciteBstWouldAddEndPuncttrue
\mciteSetBstMidEndSepPunct{\mcitedefaultmidpunct}
{\mcitedefaultendpunct}{\mcitedefaultseppunct}\relax
\EndOfBibitem
\bibitem[Lyon \emph{et~al.}(1994)Lyon, De~Leenheer, Horobin, Lambert, Schulte,
  Van~Liedekerke, and Wittekind]{lyon1994standardization}
H.~O. Lyon, A.~De~Leenheer, R.~Horobin, W.~Lambert, E.~Schulte,
  B.~Van~Liedekerke and D.~Wittekind, \emph{The Histochemical Journal}, 1994,
  \textbf{26}, 533--544\relax
\mciteBstWouldAddEndPuncttrue
\mciteSetBstMidEndSepPunct{\mcitedefaultmidpunct}
{\mcitedefaultendpunct}{\mcitedefaultseppunct}\relax
\EndOfBibitem
\bibitem[Leo \emph{et~al.}(2018)Leo, Elliott, Shih, Gupta, Feldman, and
  Madabhushi]{leo2018stable}
P.~Leo, R.~Elliott, N.~N. Shih, S.~Gupta, M.~Feldman and A.~Madabhushi,
  \emph{Scientific reports}, 2018, \textbf{8}, 1--13\relax
\mciteBstWouldAddEndPuncttrue
\mciteSetBstMidEndSepPunct{\mcitedefaultmidpunct}
{\mcitedefaultendpunct}{\mcitedefaultseppunct}\relax
\EndOfBibitem
\bibitem[Allsbrook~Jr \emph{et~al.}(2001)Allsbrook~Jr, Mangold, Johnson, Lane,
  Lane, and Epstein]{allsbrook2001interobserver}
W.~C. Allsbrook~Jr, K.~A. Mangold, M.~H. Johnson, R.~B. Lane, C.~G. Lane and
  J.~I. Epstein, \emph{Human pathology}, 2001, \textbf{32}, 81--88\relax
\mciteBstWouldAddEndPuncttrue
\mciteSetBstMidEndSepPunct{\mcitedefaultmidpunct}
{\mcitedefaultendpunct}{\mcitedefaultseppunct}\relax
\EndOfBibitem
\bibitem[Borowsky \emph{et~al.}(2020)Borowsky, Glassy, Wallace, Kallichanda,
  Behling, Miller, Oswal, Feddersen, Bakhtar,
  Mendoza,\emph{et~al.}]{borowsky2020digital}
A.~D. Borowsky, E.~F. Glassy, W.~D. Wallace, N.~S. Kallichanda, C.~A. Behling,
  D.~V. Miller, H.~N. Oswal, R.~M. Feddersen, O.~R. Bakhtar, A.~E. Mendoza
  \emph{et~al.}, \emph{Archives of Pathology \& Laboratory Medicine},
  2020\relax
\mciteBstWouldAddEndPuncttrue
\mciteSetBstMidEndSepPunct{\mcitedefaultmidpunct}
{\mcitedefaultendpunct}{\mcitedefaultseppunct}\relax
\EndOfBibitem
\bibitem[Mosquera-Lopez \emph{et~al.}(2014)Mosquera-Lopez, Agaian, Velez-Hoyos,
  and Thompson]{mosquera2014computer}
C.~Mosquera-Lopez, S.~Agaian, A.~Velez-Hoyos and I.~Thompson, \emph{IEEE
  reviews in biomedical engineering}, 2014, \textbf{8}, 98--113\relax
\mciteBstWouldAddEndPuncttrue
\mciteSetBstMidEndSepPunct{\mcitedefaultmidpunct}
{\mcitedefaultendpunct}{\mcitedefaultseppunct}\relax
\EndOfBibitem
\bibitem[Irshad \emph{et~al.}(2013)Irshad, Veillard, Roux, and
  Racoceanu]{irshad2013methods}
H.~Irshad, A.~Veillard, L.~Roux and D.~Racoceanu, \emph{IEEE reviews in
  biomedical engineering}, 2013, \textbf{7}, 97--114\relax
\mciteBstWouldAddEndPuncttrue
\mciteSetBstMidEndSepPunct{\mcitedefaultmidpunct}
{\mcitedefaultendpunct}{\mcitedefaultseppunct}\relax
\EndOfBibitem
\bibitem[Madabhushi and Lee(2016)]{madabhushi2016image}
A.~Madabhushi and G.~Lee, \emph{Image analysis and machine learning in digital
  pathology: Challenges and opportunities}, 2016\relax
\mciteBstWouldAddEndPuncttrue
\mciteSetBstMidEndSepPunct{\mcitedefaultmidpunct}
{\mcitedefaultendpunct}{\mcitedefaultseppunct}\relax
\EndOfBibitem
\bibitem[Dybas \emph{et~al.}(2016)Dybas, Marzec, Pacia, Kochan, Czamara,
  Chrabaszcz, Staniszewska-Slezak, Malek, Baranska, and Kaczor]{dybas2016raman}
J.~Dybas, K.~M. Marzec, M.~Z. Pacia, K.~Kochan, K.~Czamara, K.~Chrabaszcz,
  E.~Staniszewska-Slezak, K.~Malek, M.~Baranska and A.~Kaczor, \emph{TrAC
  Trends in Analytical Chemistry}, 2016, \textbf{85}, 117--127\relax
\mciteBstWouldAddEndPuncttrue
\mciteSetBstMidEndSepPunct{\mcitedefaultmidpunct}
{\mcitedefaultendpunct}{\mcitedefaultseppunct}\relax
\EndOfBibitem
\bibitem[Baker \emph{et~al.}(2018)Baker, Byrne, Chalmers, Gardner, Goodacre,
  Henderson, Kazarian, Martin, Moger, Stone,\emph{et~al.}]{baker2018clinical}
M.~J. Baker, H.~J. Byrne, J.~Chalmers, P.~Gardner, R.~Goodacre, A.~Henderson,
  S.~G. Kazarian, F.~L. Martin, J.~Moger, N.~Stone \emph{et~al.},
  \emph{Analyst}, 2018, \textbf{143}, 1735--1757\relax
\mciteBstWouldAddEndPuncttrue
\mciteSetBstMidEndSepPunct{\mcitedefaultmidpunct}
{\mcitedefaultendpunct}{\mcitedefaultseppunct}\relax
\EndOfBibitem
\bibitem[Krafft \emph{et~al.}(2017)Krafft, Schmitt, Schie, Cialla-May,
  Matth{\"a}us, Bocklitz, and Popp]{krafft2017label}
C.~Krafft, M.~Schmitt, I.~W. Schie, D.~Cialla-May, C.~Matth{\"a}us, T.~Bocklitz
  and J.~Popp, \emph{Angewandte Chemie International Edition}, 2017,
  \textbf{56}, 4392--4430\relax
\mciteBstWouldAddEndPuncttrue
\mciteSetBstMidEndSepPunct{\mcitedefaultmidpunct}
{\mcitedefaultendpunct}{\mcitedefaultseppunct}\relax
\EndOfBibitem
\bibitem[Roman \emph{et~al.}(2019)Roman, Wrobel, Panek, Efeoglu,
  Wiltowska-Zuber, Paluszkiewicz, Byrne, and Kwiatek]{roman2019exploring}
M.~Roman, T.~P. Wrobel, A.~Panek, E.~Efeoglu, J.~Wiltowska-Zuber,
  C.~Paluszkiewicz, H.~J. Byrne and W.~M. Kwiatek, \emph{Scientific reports},
  2019, \textbf{9}, 1--13\relax
\mciteBstWouldAddEndPuncttrue
\mciteSetBstMidEndSepPunct{\mcitedefaultmidpunct}
{\mcitedefaultendpunct}{\mcitedefaultseppunct}\relax
\EndOfBibitem
\bibitem[Movasaghi \emph{et~al.}(2007)Movasaghi, Rehman, and
  Rehman]{movasaghi2007raman}
Z.~Movasaghi, S.~Rehman and I.~U. Rehman, \emph{Applied Spectroscopy Reviews},
  2007, \textbf{42}, 493--541\relax
\mciteBstWouldAddEndPuncttrue
\mciteSetBstMidEndSepPunct{\mcitedefaultmidpunct}
{\mcitedefaultendpunct}{\mcitedefaultseppunct}\relax
\EndOfBibitem
\bibitem[Chowdary \emph{et~al.}(2007)Chowdary, Kumar, Thakur, Anand, Kurien,
  Krishna, and Mathew]{chowdary2007discrimination}
M.~Chowdary, K.~K. Kumar, K.~Thakur, A.~Anand, J.~Kurien, C.~M. Krishna and
  S.~Mathew, \emph{Photomedicine and laser surgery}, 2007, \textbf{25},
  269--274\relax
\mciteBstWouldAddEndPuncttrue
\mciteSetBstMidEndSepPunct{\mcitedefaultmidpunct}
{\mcitedefaultendpunct}{\mcitedefaultseppunct}\relax
\EndOfBibitem
\bibitem[Brozek-Pluska \emph{et~al.}(2019)Brozek-Pluska, Musial, Kordek, and
  Abramczyk]{brozek2019analysis}
B.~Brozek-Pluska, J.~Musial, R.~Kordek and H.~Abramczyk, \emph{International
  journal of molecular sciences}, 2019, \textbf{20}, 3398\relax
\mciteBstWouldAddEndPuncttrue
\mciteSetBstMidEndSepPunct{\mcitedefaultmidpunct}
{\mcitedefaultendpunct}{\mcitedefaultseppunct}\relax
\EndOfBibitem
\bibitem[Gniadecka \emph{et~al.}(2004)Gniadecka, Philipsen, Wessel, Gniadecki,
  Wulf, Sigurdsson, Nielsen, Christensen, Hercogova,
  Rossen,\emph{et~al.}]{gniadecka2004melanoma}
M.~Gniadecka, P.~A. Philipsen, S.~Wessel, R.~Gniadecki, H.~C. Wulf,
  S.~Sigurdsson, O.~F. Nielsen, D.~H. Christensen, J.~Hercogova, K.~Rossen
  \emph{et~al.}, \emph{Journal of investigative dermatology}, 2004,
  \textbf{122}, 443--449\relax
\mciteBstWouldAddEndPuncttrue
\mciteSetBstMidEndSepPunct{\mcitedefaultmidpunct}
{\mcitedefaultendpunct}{\mcitedefaultseppunct}\relax
\EndOfBibitem
\bibitem[Haka \emph{et~al.}(2005)Haka, Shafer-Peltier, Fitzmaurice, Crowe,
  Dasari, and Feld]{haka2005diagnosing}
A.~S. Haka, K.~E. Shafer-Peltier, M.~Fitzmaurice, J.~Crowe, R.~R. Dasari and
  M.~S. Feld, \emph{Proceedings of the National Academy of Sciences}, 2005,
  \textbf{102}, 12371--12376\relax
\mciteBstWouldAddEndPuncttrue
\mciteSetBstMidEndSepPunct{\mcitedefaultmidpunct}
{\mcitedefaultendpunct}{\mcitedefaultseppunct}\relax
\EndOfBibitem
\bibitem[Huang \emph{et~al.}(2003)Huang, McWilliams, Lui, McLean, Lam, and
  Zeng]{huang2003near}
Z.~Huang, A.~McWilliams, H.~Lui, D.~I. McLean, S.~Lam and H.~Zeng,
  \emph{International journal of cancer}, 2003, \textbf{107}, 1047--1052\relax
\mciteBstWouldAddEndPuncttrue
\mciteSetBstMidEndSepPunct{\mcitedefaultmidpunct}
{\mcitedefaultendpunct}{\mcitedefaultseppunct}\relax
\EndOfBibitem
\bibitem[Lyng \emph{et~al.}(2007)Lyng, Faol{\'a}in, Conroy, Meade, Knief,
  Duffy, Hunter, Byrne, Kelehan, and Byrne]{lyng2007vibrational}
F.~M. Lyng, E.~{\'O}. Faol{\'a}in, J.~Conroy, A.~Meade, P.~Knief, B.~Duffy,
  M.~Hunter, J.~Byrne, P.~Kelehan and H.~Byrne, \emph{Experimental and
  molecular pathology}, 2007, \textbf{82}, 121--129\relax
\mciteBstWouldAddEndPuncttrue
\mciteSetBstMidEndSepPunct{\mcitedefaultmidpunct}
{\mcitedefaultendpunct}{\mcitedefaultseppunct}\relax
\EndOfBibitem
\bibitem[Crow \emph{et~al.}(2003)Crow, Stone, Kendall, Uff, Farmer, Barr, and
  Wright]{crow2003use}
P.~Crow, N.~Stone, C.~Kendall, J.~Uff, J.~Farmer, H.~Barr and M.~Wright,
  \emph{British journal of cancer}, 2003, \textbf{89}, 106--108\relax
\mciteBstWouldAddEndPuncttrue
\mciteSetBstMidEndSepPunct{\mcitedefaultmidpunct}
{\mcitedefaultendpunct}{\mcitedefaultseppunct}\relax
\EndOfBibitem
\bibitem[Devpura \emph{et~al.}(2010)Devpura, Thakur, Sarkar, Sakr, Naik, and
  Naik]{devpura2010detection}
S.~Devpura, J.~S. Thakur, F.~H. Sarkar, W.~A. Sakr, V.~M. Naik and R.~Naik,
  \emph{Vibrational Spectroscopy}, 2010, \textbf{53}, 227--232\relax
\mciteBstWouldAddEndPuncttrue
\mciteSetBstMidEndSepPunct{\mcitedefaultmidpunct}
{\mcitedefaultendpunct}{\mcitedefaultseppunct}\relax
\EndOfBibitem
\bibitem[Tollefson \emph{et~al.}(2010)Tollefson, Magera, Sebo, Cohen, Drauch,
  Maier, and Frank]{tollefson2010raman}
M.~Tollefson, J.~Magera, T.~Sebo, J.~Cohen, A.~Drauch, J.~Maier and I.~Frank,
  \emph{BJU international}, 2010, \textbf{106}, 484--488\relax
\mciteBstWouldAddEndPuncttrue
\mciteSetBstMidEndSepPunct{\mcitedefaultmidpunct}
{\mcitedefaultendpunct}{\mcitedefaultseppunct}\relax
\EndOfBibitem
\bibitem[Kast \emph{et~al.}(2014)Kast, Tucker, Killian, Trexler, Honn, and
  Auner]{kast2014emerging}
R.~E. Kast, S.~C. Tucker, K.~Killian, M.~Trexler, K.~V. Honn and G.~W. Auner,
  \emph{Cancer and Metastasis Reviews}, 2014, \textbf{33}, 673--693\relax
\mciteBstWouldAddEndPuncttrue
\mciteSetBstMidEndSepPunct{\mcitedefaultmidpunct}
{\mcitedefaultendpunct}{\mcitedefaultseppunct}\relax
\EndOfBibitem
\bibitem[Samiei \emph{et~al.}(2018)Samiei, Miller, Cohen, Gomer, Stewart, and
  Treado]{samiei2018prospective}
A.~Samiei, R.~Miller, J.~Cohen, H.~Gomer, S.~Stewart and P.~Treado, CANCER
  RESEARCH, 2018, pp. 121--122\relax
\mciteBstWouldAddEndPuncttrue
\mciteSetBstMidEndSepPunct{\mcitedefaultmidpunct}
{\mcitedefaultendpunct}{\mcitedefaultseppunct}\relax
\EndOfBibitem
\bibitem[Crow \emph{et~al.}(2005)Crow, Molckovsky, Stone, Uff, Wilson, and
  WongKeeSong]{crow2005assessment}
P.~Crow, A.~Molckovsky, N.~Stone, J.~Uff, B.~Wilson and L.-M. WongKeeSong,
  \emph{Urology}, 2005, \textbf{65}, 1126--1130\relax
\mciteBstWouldAddEndPuncttrue
\mciteSetBstMidEndSepPunct{\mcitedefaultmidpunct}
{\mcitedefaultendpunct}{\mcitedefaultseppunct}\relax
\EndOfBibitem
\bibitem[Li \emph{et~al.}(2014)Li, Zhang, Xu, Li, Zeng, Lin, Guo, Liu, Xiong,
  and Liu]{li2014noninvasive}
S.~Li, Y.~Zhang, J.~Xu, L.~Li, Q.~Zeng, L.~Lin, Z.~Guo, Z.~Liu, H.~Xiong and
  S.~Liu, \emph{Applied Physics Letters}, 2014, \textbf{105}, 091104\relax
\mciteBstWouldAddEndPuncttrue
\mciteSetBstMidEndSepPunct{\mcitedefaultmidpunct}
{\mcitedefaultendpunct}{\mcitedefaultseppunct}\relax
\EndOfBibitem
\bibitem[Wang \emph{et~al.}(2013)Wang, He, Zeng, Fuan, Dang, Wang, Wang, Huang,
  Cao, Zhang,\emph{et~al.}]{wang2013raman}
L.~Wang, D.~He, J.~Zeng, Z.~Fuan, Q.~Dang, X.~Wang, J.~Wang, L.~Huang, P.~Cao,
  G.~Zhang \emph{et~al.}, \emph{Journal of biomedical optics}, 2013,
  \textbf{18}, 087001\relax
\mciteBstWouldAddEndPuncttrue
\mciteSetBstMidEndSepPunct{\mcitedefaultmidpunct}
{\mcitedefaultendpunct}{\mcitedefaultseppunct}\relax
\EndOfBibitem
\bibitem[Crow \emph{et~al.}(2005)Crow, Barrass, Kendall, Hart-Prieto, Wright,
  Persad, and Stone]{crow2005use}
P.~Crow, B.~Barrass, C.~Kendall, M.~Hart-Prieto, M.~Wright, R.~Persad and
  N.~Stone, \emph{British journal of cancer}, 2005, \textbf{92},
  2166--2170\relax
\mciteBstWouldAddEndPuncttrue
\mciteSetBstMidEndSepPunct{\mcitedefaultmidpunct}
{\mcitedefaultendpunct}{\mcitedefaultseppunct}\relax
\EndOfBibitem
\bibitem[Del~Mistro \emph{et~al.}(2015)Del~Mistro, Cervo, Mansutti, Spizzo,
  Colombatti, Belmonte, Zucconelli, Steffan, Sergo, and
  Bonifacio]{del2015surface}
G.~Del~Mistro, S.~Cervo, E.~Mansutti, R.~Spizzo, A.~Colombatti, P.~Belmonte,
  R.~Zucconelli, A.~Steffan, V.~Sergo and A.~Bonifacio, \emph{Analytical and
  Bioanalytical Chemistry}, 2015, \textbf{407}, 3271--3275\relax
\mciteBstWouldAddEndPuncttrue
\mciteSetBstMidEndSepPunct{\mcitedefaultmidpunct}
{\mcitedefaultendpunct}{\mcitedefaultseppunct}\relax
\EndOfBibitem
\bibitem[Du \emph{et~al.}(2016)Du, Li, Lu, and Xiao]{du2016overview}
J.~Du, W.~Li, K.~Lu and B.~Xiao, \emph{Neurocomputing}, 2016, \textbf{215},
  3--20\relax
\mciteBstWouldAddEndPuncttrue
\mciteSetBstMidEndSepPunct{\mcitedefaultmidpunct}
{\mcitedefaultendpunct}{\mcitedefaultseppunct}\relax
\EndOfBibitem
\bibitem[James and Dasarathy(2014)]{james2014medical}
A.~P. James and B.~V. Dasarathy, \emph{Information fusion}, 2014, \textbf{19},
  4--19\relax
\mciteBstWouldAddEndPuncttrue
\mciteSetBstMidEndSepPunct{\mcitedefaultmidpunct}
{\mcitedefaultendpunct}{\mcitedefaultseppunct}\relax
\EndOfBibitem
\bibitem[Kwak \emph{et~al.}(2011)Kwak, Hewitt, Sinha, and
  Bhargava]{kwak2011multimodal}
J.~T. Kwak, S.~M. Hewitt, S.~Sinha and R.~Bhargava, \emph{BMC cancer}, 2011,
  \textbf{11}, 62\relax
\mciteBstWouldAddEndPuncttrue
\mciteSetBstMidEndSepPunct{\mcitedefaultmidpunct}
{\mcitedefaultendpunct}{\mcitedefaultseppunct}\relax
\EndOfBibitem
\bibitem[Rodner \emph{et~al.}(2019)Rodner, Bocklitz, von Eggeling, Ernst,
  Chernavskaia, Popp, Denzler, and Guntinas-Lichius]{rodner2019fully}
E.~Rodner, T.~Bocklitz, F.~von Eggeling, G.~Ernst, O.~Chernavskaia, J.~Popp,
  J.~Denzler and O.~Guntinas-Lichius, \emph{Head \& Neck}, 2019, \textbf{41},
  116--121\relax
\mciteBstWouldAddEndPuncttrue
\mciteSetBstMidEndSepPunct{\mcitedefaultmidpunct}
{\mcitedefaultendpunct}{\mcitedefaultseppunct}\relax
\EndOfBibitem
\bibitem[Patil \emph{et~al.}(2008)Patil, Bosschaart, Keller, van Leeuwen, and
  Mahadevan-Jansen]{patil2008combined}
C.~A. Patil, N.~Bosschaart, M.~D. Keller, T.~G. van Leeuwen and
  A.~Mahadevan-Jansen, \emph{Optics letters}, 2008, \textbf{33},
  1135--1137\relax
\mciteBstWouldAddEndPuncttrue
\mciteSetBstMidEndSepPunct{\mcitedefaultmidpunct}
{\mcitedefaultendpunct}{\mcitedefaultseppunct}\relax
\EndOfBibitem
\bibitem[Ashok \emph{et~al.}(2013)Ashok, Praveen, Bellini, Riches, Dholakia,
  and Herrington]{ashok2013multi}
P.~C. Ashok, B.~B. Praveen, N.~Bellini, A.~Riches, K.~Dholakia and C.~S.
  Herrington, \emph{Biomedical optics express}, 2013, \textbf{4},
  2179--2186\relax
\mciteBstWouldAddEndPuncttrue
\mciteSetBstMidEndSepPunct{\mcitedefaultmidpunct}
{\mcitedefaultendpunct}{\mcitedefaultseppunct}\relax
\EndOfBibitem
\bibitem[Yuan \emph{et~al.}(2012)Yuan, Failmezger, Rueda, Ali, Gr{\"a}f, Chin,
  Schwarz, Curtis, Dunning, Bardwell,\emph{et~al.}]{yuan2012quantitative}
Y.~Yuan, H.~Failmezger, O.~M. Rueda, H.~R. Ali, S.~Gr{\"a}f, S.-F. Chin, R.~F.
  Schwarz, C.~Curtis, M.~J. Dunning, H.~Bardwell \emph{et~al.}, \emph{Science
  translational medicine}, 2012, \textbf{4}, 157ra143--157ra143\relax
\mciteBstWouldAddEndPuncttrue
\mciteSetBstMidEndSepPunct{\mcitedefaultmidpunct}
{\mcitedefaultendpunct}{\mcitedefaultseppunct}\relax
\EndOfBibitem
\bibitem[Nguyen \emph{et~al.}(2017)Nguyen, Sridharan, Macias, Kajdacsy-Balla,
  Melamed, Do, and Popescu]{nguyen2017automatic}
T.~H. Nguyen, S.~Sridharan, V.~Macias, A.~Kajdacsy-Balla, J.~Melamed, M.~N. Do
  and G.~Popescu, \emph{Journal of biomedical optics}, 2017, \textbf{22},
  036015\relax
\mciteBstWouldAddEndPuncttrue
\mciteSetBstMidEndSepPunct{\mcitedefaultmidpunct}
{\mcitedefaultendpunct}{\mcitedefaultseppunct}\relax
\EndOfBibitem
\bibitem[Lee \emph{et~al.}(2014)Lee, Sparks, Ali, Shih, Feldman, Spangler,
  Rebbeck, Tomaszewski, and Madabhushi]{lee2014co}
G.~Lee, R.~Sparks, S.~Ali, N.~N. Shih, M.~D. Feldman, E.~Spangler, T.~Rebbeck,
  J.~E. Tomaszewski and A.~Madabhushi, \emph{PloS one}, 2014, \textbf{9},
  e97954\relax
\mciteBstWouldAddEndPuncttrue
\mciteSetBstMidEndSepPunct{\mcitedefaultmidpunct}
{\mcitedefaultendpunct}{\mcitedefaultseppunct}\relax
\EndOfBibitem
\bibitem[Veta \emph{et~al.}(2012)Veta, Kornegoor, Huisman, Verschuur-Maes,
  Viergever, Pluim, and Van~Diest]{veta2012prognostic}
M.~Veta, R.~Kornegoor, A.~Huisman, A.~H. Verschuur-Maes, M.~A. Viergever, J.~P.
  Pluim and P.~J. Van~Diest, \emph{Modern pathology}, 2012, \textbf{25},
  1559--1565\relax
\mciteBstWouldAddEndPuncttrue
\mciteSetBstMidEndSepPunct{\mcitedefaultmidpunct}
{\mcitedefaultendpunct}{\mcitedefaultseppunct}\relax
\EndOfBibitem
\bibitem[Doyle \emph{et~al.}(2007)Doyle, Hwang, Shah, Madabhushi, Feldman, and
  Tomaszeweski]{doyle2007automated}
S.~Doyle, M.~Hwang, K.~Shah, A.~Madabhushi, M.~Feldman and J.~Tomaszeweski,
  2007 4th IEEE International Symposium on Biomedical Imaging: From Nano to
  Macro, 2007, pp. 1284--1287\relax
\mciteBstWouldAddEndPuncttrue
\mciteSetBstMidEndSepPunct{\mcitedefaultmidpunct}
{\mcitedefaultendpunct}{\mcitedefaultseppunct}\relax
\EndOfBibitem
\bibitem[Sparks and Madabhushi(2013)]{sparks2013explicit}
R.~Sparks and A.~Madabhushi, \emph{Medical image analysis}, 2013, \textbf{17},
  997--1009\relax
\mciteBstWouldAddEndPuncttrue
\mciteSetBstMidEndSepPunct{\mcitedefaultmidpunct}
{\mcitedefaultendpunct}{\mcitedefaultseppunct}\relax
\EndOfBibitem
\bibitem[Linder \emph{et~al.}(2012)Linder, Konsti, Turkki, Rahtu, Lundin,
  Nordling, Haglund, Ahonen, Pietik{\"a}inen, and
  Lundin]{linder2012identification}
N.~Linder, J.~Konsti, R.~Turkki, E.~Rahtu, M.~Lundin, S.~Nordling, C.~Haglund,
  T.~Ahonen, M.~Pietik{\"a}inen and J.~Lundin, \emph{Diagnostic pathology},
  2012, \textbf{7}, 22\relax
\mciteBstWouldAddEndPuncttrue
\mciteSetBstMidEndSepPunct{\mcitedefaultmidpunct}
{\mcitedefaultendpunct}{\mcitedefaultseppunct}\relax
\EndOfBibitem
\bibitem[Jafari-Khouzani and Soltanian-Zadeh(2003)]{jafari2003multiwavelet}
K.~Jafari-Khouzani and H.~Soltanian-Zadeh, \emph{IEEE Transactions on
  Biomedical Engineering}, 2003, \textbf{50}, 697--704\relax
\mciteBstWouldAddEndPuncttrue
\mciteSetBstMidEndSepPunct{\mcitedefaultmidpunct}
{\mcitedefaultendpunct}{\mcitedefaultseppunct}\relax
\EndOfBibitem
\bibitem[Sanghavi and Agaian(2016)]{sanghavi2016automated}
F.~M. Sanghavi and S.~S. Agaian, Mobile Multimedia/Image Processing, Security,
  and Applications 2016, 2016, p. 98690T\relax
\mciteBstWouldAddEndPuncttrue
\mciteSetBstMidEndSepPunct{\mcitedefaultmidpunct}
{\mcitedefaultendpunct}{\mcitedefaultseppunct}\relax
\EndOfBibitem
\bibitem[Lowe(2004)]{lowe2004distinctive}
D.~G. Lowe, \emph{International journal of computer vision}, 2004, \textbf{60},
  91--110\relax
\mciteBstWouldAddEndPuncttrue
\mciteSetBstMidEndSepPunct{\mcitedefaultmidpunct}
{\mcitedefaultendpunct}{\mcitedefaultseppunct}\relax
\EndOfBibitem
\bibitem[Bychkov \emph{et~al.}(2018)Bychkov, Linder, Turkki, Nordling, Kovanen,
  Verrill, Walliander, Lundin, Haglund, and Lundin]{bychkov2018deep}
D.~Bychkov, N.~Linder, R.~Turkki, S.~Nordling, P.~E. Kovanen, C.~Verrill,
  M.~Walliander, M.~Lundin, C.~Haglund and J.~Lundin, \emph{Scientific
  reports}, 2018, \textbf{8}, 1--11\relax
\mciteBstWouldAddEndPuncttrue
\mciteSetBstMidEndSepPunct{\mcitedefaultmidpunct}
{\mcitedefaultendpunct}{\mcitedefaultseppunct}\relax
\EndOfBibitem
\bibitem[Ertosun and Rubin(2015)]{ertosun2015automated}
M.~G. Ertosun and D.~L. Rubin, AMIA Annual Symposium Proceedings, 2015, p.
  1899\relax
\mciteBstWouldAddEndPuncttrue
\mciteSetBstMidEndSepPunct{\mcitedefaultmidpunct}
{\mcitedefaultendpunct}{\mcitedefaultseppunct}\relax
\EndOfBibitem
\bibitem[Shie \emph{et~al.}(2015)Shie, Chuang, Chou, Wu, and
  Chang]{shie2015transfer}
C.-K. Shie, C.-H. Chuang, C.-N. Chou, M.-H. Wu and E.~Y. Chang, 2015 37th
  annual international conference of the IEEE Engineering in Medicine and
  Biology Society (EMBC), 2015, pp. 711--714\relax
\mciteBstWouldAddEndPuncttrue
\mciteSetBstMidEndSepPunct{\mcitedefaultmidpunct}
{\mcitedefaultendpunct}{\mcitedefaultseppunct}\relax
\EndOfBibitem
\bibitem[Weng \emph{et~al.}(2017)Weng, Xu, Li, and Wong]{weng2017combining}
S.~Weng, X.~Xu, J.~Li and S.~T. Wong, \emph{Journal of biomedical optics},
  2017, \textbf{22}, 106017\relax
\mciteBstWouldAddEndPuncttrue
\mciteSetBstMidEndSepPunct{\mcitedefaultmidpunct}
{\mcitedefaultendpunct}{\mcitedefaultseppunct}\relax
\EndOfBibitem
\bibitem[Esteva \emph{et~al.}(2017)Esteva, Kuprel, Novoa, Ko, Swetter, Blau,
  and Thrun]{esteva2017dermatologist}
A.~Esteva, B.~Kuprel, R.~A. Novoa, J.~Ko, S.~M. Swetter, H.~M. Blau and
  S.~Thrun, \emph{nature}, 2017, \textbf{542}, 115--118\relax
\mciteBstWouldAddEndPuncttrue
\mciteSetBstMidEndSepPunct{\mcitedefaultmidpunct}
{\mcitedefaultendpunct}{\mcitedefaultseppunct}\relax
\EndOfBibitem
\bibitem[Ribeiro \emph{et~al.}(2016)Ribeiro, Uhl, Wimmer, and
  H{\"a}fner]{ribeiro2016exploring}
E.~Ribeiro, A.~Uhl, G.~Wimmer and M.~H{\"a}fner, \emph{Computational and
  mathematical methods in medicine}, 2016, \textbf{2016}, year\relax
\mciteBstWouldAddEndPuncttrue
\mciteSetBstMidEndSepPunct{\mcitedefaultmidpunct}
{\mcitedefaultendpunct}{\mcitedefaultseppunct}\relax
\EndOfBibitem
\bibitem[Chi \emph{et~al.}(2017)Chi, Walia, Babyn, Wang, Groot, and
  Eramian]{chi2017thyroid}
J.~Chi, E.~Walia, P.~Babyn, J.~Wang, G.~Groot and M.~Eramian, \emph{Journal of
  digital imaging}, 2017, \textbf{30}, 477--486\relax
\mciteBstWouldAddEndPuncttrue
\mciteSetBstMidEndSepPunct{\mcitedefaultmidpunct}
{\mcitedefaultendpunct}{\mcitedefaultseppunct}\relax
\EndOfBibitem
\bibitem[Shin \emph{et~al.}(2016)Shin, Roth, Gao, Lu, Xu, Nogues, Yao, Mollura,
  and Summers]{shin2016deep}
H.-C. Shin, H.~R. Roth, M.~Gao, L.~Lu, Z.~Xu, I.~Nogues, J.~Yao, D.~Mollura and
  R.~M. Summers, \emph{IEEE transactions on medical imaging}, 2016,
  \textbf{35}, 1285--1298\relax
\mciteBstWouldAddEndPuncttrue
\mciteSetBstMidEndSepPunct{\mcitedefaultmidpunct}
{\mcitedefaultendpunct}{\mcitedefaultseppunct}\relax
\EndOfBibitem
\bibitem[Ishioka \emph{et~al.}(2018)Ishioka, Matsuoka, Uehara, Yasuda, Kijima,
  Yoshida, Yokoyama, Saito, Kihara, Numao,\emph{et~al.}]{ishioka2018computer}
J.~Ishioka, Y.~Matsuoka, S.~Uehara, Y.~Yasuda, T.~Kijima, S.~Yoshida,
  M.~Yokoyama, K.~Saito, K.~Kihara, N.~Numao \emph{et~al.}, \emph{BJU
  international}, 2018, \textbf{122}, 411--417\relax
\mciteBstWouldAddEndPuncttrue
\mciteSetBstMidEndSepPunct{\mcitedefaultmidpunct}
{\mcitedefaultendpunct}{\mcitedefaultseppunct}\relax
\EndOfBibitem
\bibitem[Arvaniti \emph{et~al.}(2018)Arvaniti, Fricker, Moret, Rupp, Hermanns,
  Fankhauser, Wey, Wild, Rueschoff, and Claassen]{arvaniti2018automated}
E.~Arvaniti, K.~S. Fricker, M.~Moret, N.~Rupp, T.~Hermanns, C.~Fankhauser,
  N.~Wey, P.~J. Wild, J.~H. Rueschoff and M.~Claassen, \emph{Scientific
  reports}, 2018, \textbf{8}, 1--11\relax
\mciteBstWouldAddEndPuncttrue
\mciteSetBstMidEndSepPunct{\mcitedefaultmidpunct}
{\mcitedefaultendpunct}{\mcitedefaultseppunct}\relax
\EndOfBibitem
\bibitem[Litjens \emph{et~al.}(2016)Litjens, S{\'a}nchez, Timofeeva, Hermsen,
  Nagtegaal, Kovacs, Hulsbergen-Van De~Kaa, Bult, Van~Ginneken, and Van
  Der~Laak]{litjens2016deep}
G.~Litjens, C.~I. S{\'a}nchez, N.~Timofeeva, M.~Hermsen, I.~Nagtegaal,
  I.~Kovacs, C.~Hulsbergen-Van De~Kaa, P.~Bult, B.~Van~Ginneken and J.~Van
  Der~Laak, \emph{Scientific reports}, 2016, \textbf{6}, 26286\relax
\mciteBstWouldAddEndPuncttrue
\mciteSetBstMidEndSepPunct{\mcitedefaultmidpunct}
{\mcitedefaultendpunct}{\mcitedefaultseppunct}\relax
\EndOfBibitem
\bibitem[Song \emph{et~al.}(2018)Song, Zhang, Yan, Liu, Zhou, Hu, and
  Yang]{song2018computer}
Y.~Song, Y.-D. Zhang, X.~Yan, H.~Liu, M.~Zhou, B.~Hu and G.~Yang, \emph{Journal
  of Magnetic Resonance Imaging}, 2018, \textbf{48}, 1570--1577\relax
\mciteBstWouldAddEndPuncttrue
\mciteSetBstMidEndSepPunct{\mcitedefaultmidpunct}
{\mcitedefaultendpunct}{\mcitedefaultseppunct}\relax
\EndOfBibitem
\bibitem[O’Mahony \emph{et~al.}(2019)O’Mahony, Campbell, Carvalho,
  Harapanahalli, Hernandez, Krpalkova, Riordan, and Walsh]{o2019deep}
N.~O’Mahony, S.~Campbell, A.~Carvalho, S.~Harapanahalli, G.~V. Hernandez,
  L.~Krpalkova, D.~Riordan and J.~Walsh, Science and Information Conference,
  2019, pp. 128--144\relax
\mciteBstWouldAddEndPuncttrue
\mciteSetBstMidEndSepPunct{\mcitedefaultmidpunct}
{\mcitedefaultendpunct}{\mcitedefaultseppunct}\relax
\EndOfBibitem
\bibitem[Wang \emph{et~al.}(2017)Wang, Yang, Weinreb, Han, Li, Kong, Yan, Ke,
  Luo, Liu,\emph{et~al.}]{wang2017searching}
X.~Wang, W.~Yang, J.~Weinreb, J.~Han, Q.~Li, X.~Kong, Y.~Yan, Z.~Ke, B.~Luo,
  T.~Liu \emph{et~al.}, \emph{Scientific reports}, 2017, \textbf{7}, 1--8\relax
\mciteBstWouldAddEndPuncttrue
\mciteSetBstMidEndSepPunct{\mcitedefaultmidpunct}
{\mcitedefaultendpunct}{\mcitedefaultseppunct}\relax
\EndOfBibitem
\bibitem[Russakovsky \emph{et~al.}(2015)Russakovsky, Deng, Su, Krause,
  Satheesh, Ma, Huang, Karpathy, Khosla,
  Bernstein,\emph{et~al.}]{russakovsky2015imagenet}
O.~Russakovsky, J.~Deng, H.~Su, J.~Krause, S.~Satheesh, S.~Ma, Z.~Huang,
  A.~Karpathy, A.~Khosla, M.~Bernstein \emph{et~al.}, \emph{International
  journal of computer vision}, 2015, \textbf{115}, 211--252\relax
\mciteBstWouldAddEndPuncttrue
\mciteSetBstMidEndSepPunct{\mcitedefaultmidpunct}
{\mcitedefaultendpunct}{\mcitedefaultseppunct}\relax
\EndOfBibitem
\bibitem[Breen \emph{et~al.}(2017)Breen, O'Neill, Murphy, Fan, Boyce,
  Fitzgerald, Dorris, Brady, Finn, Hayes,\emph{et~al.}]{breen2017investigating}
K.~J. Breen, A.~O'Neill, L.~Murphy, Y.~Fan, S.~Boyce, N.~Fitzgerald, E.~Dorris,
  L.~Brady, S.~P. Finn, B.~D. Hayes \emph{et~al.}, \emph{The Prostate}, 2017,
  \textbf{77}, 1288--1300\relax
\mciteBstWouldAddEndPuncttrue
\mciteSetBstMidEndSepPunct{\mcitedefaultmidpunct}
{\mcitedefaultendpunct}{\mcitedefaultseppunct}\relax
\EndOfBibitem
\bibitem[Hospital()]{stmichaels}
S.~M. Hospital, http://www.stmichaelshospital.com/research
  /facilities/images/histology-methods-hematoxylin-eosin-staining-manual-protocol.pdf
  (Hematoxylin and Eosin (H\&E) Staining – Manual Protocol)\relax
\mciteBstWouldAddEndPuncttrue
\mciteSetBstMidEndSepPunct{\mcitedefaultmidpunct}
{\mcitedefaultendpunct}{\mcitedefaultseppunct}\relax
\EndOfBibitem
\bibitem[Rahman \emph{et~al.}(2020)Rahman, Jahangir, Lynch, Alattar, Aura,
  Russell, Lanigan, and Gallagher]{rahman2020advances}
A.~Rahman, C.~Jahangir, S.~M. Lynch, N.~Alattar, C.~Aura, N.~Russell,
  F.~Lanigan and W.~M. Gallagher, \emph{Expert Review of Molecular
  Diagnostics}, 2020,  1--11\relax
\mciteBstWouldAddEndPuncttrue
\mciteSetBstMidEndSepPunct{\mcitedefaultmidpunct}
{\mcitedefaultendpunct}{\mcitedefaultseppunct}\relax
\EndOfBibitem
\bibitem[Lindeberg(2012)]{lindeberg2012scale}
T.~Lindeberg, 2012\relax
\mciteBstWouldAddEndPuncttrue
\mciteSetBstMidEndSepPunct{\mcitedefaultmidpunct}
{\mcitedefaultendpunct}{\mcitedefaultseppunct}\relax
\EndOfBibitem
\bibitem[Csurka \emph{et~al.}(2004)Csurka, Dance, Fan, Willamowski, and
  Bray]{csurka2004visual}
G.~Csurka, C.~Dance, L.~Fan, J.~Willamowski and C.~Bray, Workshop on
  statistical learning in computer vision, ECCV, 2004, pp. 1--2\relax
\mciteBstWouldAddEndPuncttrue
\mciteSetBstMidEndSepPunct{\mcitedefaultmidpunct}
{\mcitedefaultendpunct}{\mcitedefaultseppunct}\relax
\EndOfBibitem
\bibitem[Csurka and Humenberger(2018)]{csurka2018handcrafted}
G.~Csurka and M.~Humenberger, \emph{arXiv preprint arXiv:1807.10254}, 2018,
  \textbf{2}, year\relax
\mciteBstWouldAddEndPuncttrue
\mciteSetBstMidEndSepPunct{\mcitedefaultmidpunct}
{\mcitedefaultendpunct}{\mcitedefaultseppunct}\relax
\EndOfBibitem
\bibitem[Yang \emph{et~al.}(2007)Yang, Jiang, Hauptmann, and
  Ngo]{yang2007evaluating}
J.~Yang, Y.-G. Jiang, A.~G. Hauptmann and C.-W. Ngo, Proceedings of the
  international workshop on Workshop on multimedia information retrieval, 2007,
  pp. 197--206\relax
\mciteBstWouldAddEndPuncttrue
\mciteSetBstMidEndSepPunct{\mcitedefaultmidpunct}
{\mcitedefaultendpunct}{\mcitedefaultseppunct}\relax
\EndOfBibitem
\bibitem[Lin \emph{et~al.}(2019)Lin, Lin, Cao, Velmurugan, Ward, and
  Ober]{lin2019two}
D.~Lin, Z.~Lin, J.~Cao, R.~Velmurugan, E.~S. Ward and R.~J. Ober, \emph{PloS
  one}, 2019, \textbf{14}, e0218931\relax
\mciteBstWouldAddEndPuncttrue
\mciteSetBstMidEndSepPunct{\mcitedefaultmidpunct}
{\mcitedefaultendpunct}{\mcitedefaultseppunct}\relax
\EndOfBibitem
\bibitem[Suh \emph{et~al.}(2018)Suh, Hofstee, IJsselmuiden, and van
  Henten]{suh2018sugar}
H.~K. Suh, J.~W. Hofstee, J.~IJsselmuiden and E.~J. van Henten,
  \emph{Biosystems Engineering}, 2018, \textbf{166}, 210--226\relax
\mciteBstWouldAddEndPuncttrue
\mciteSetBstMidEndSepPunct{\mcitedefaultmidpunct}
{\mcitedefaultendpunct}{\mcitedefaultseppunct}\relax
\EndOfBibitem
\bibitem[Pedregosa \emph{et~al.}(2011)Pedregosa, Varoquaux, Gramfort, Michel,
  Thirion, Grisel, Blondel, Prettenhofer, Weiss,
  Dubourg,\emph{et~al.}]{pedregosa2011scikit}
F.~Pedregosa, G.~Varoquaux, A.~Gramfort, V.~Michel, B.~Thirion, O.~Grisel,
  M.~Blondel, P.~Prettenhofer, R.~Weiss, V.~Dubourg \emph{et~al.}, \emph{the
  Journal of machine Learning research}, 2011, \textbf{12}, 2825--2830\relax
\mciteBstWouldAddEndPuncttrue
\mciteSetBstMidEndSepPunct{\mcitedefaultmidpunct}
{\mcitedefaultendpunct}{\mcitedefaultseppunct}\relax
\EndOfBibitem
\bibitem[Hsu \emph{et~al.}(2003)Hsu, Chang,
  Lin,\emph{et~al.}]{hsu2003practical}
C.-W. Hsu, C.-C. Chang, C.-J. Lin \emph{et~al.}, \emph{A practical guide to
  support vector classification}, 2003\relax
\mciteBstWouldAddEndPuncttrue
\mciteSetBstMidEndSepPunct{\mcitedefaultmidpunct}
{\mcitedefaultendpunct}{\mcitedefaultseppunct}\relax
\EndOfBibitem
\bibitem[Dorrepaal \emph{et~al.}(2016)Dorrepaal, Malegori, and
  Gowen]{dorrepaal2016tutorial}
R.~Dorrepaal, C.~Malegori and A.~Gowen, \emph{Journal of Near Infrared
  Spectroscopy}, 2016, \textbf{24}, 89--107\relax
\mciteBstWouldAddEndPuncttrue
\mciteSetBstMidEndSepPunct{\mcitedefaultmidpunct}
{\mcitedefaultendpunct}{\mcitedefaultseppunct}\relax
\EndOfBibitem
\bibitem[Das \emph{et~al.}(2017)Das, Dai, Liu, Hu, Tong, Chen, and
  Smith]{das2017raman}
N.~K. Das, Y.~Dai, P.~Liu, C.~Hu, L.~Tong, X.~Chen and Z.~J. Smith,
  \emph{Sensors}, 2017, \textbf{17}, 1592\relax
\mciteBstWouldAddEndPuncttrue
\mciteSetBstMidEndSepPunct{\mcitedefaultmidpunct}
{\mcitedefaultendpunct}{\mcitedefaultseppunct}\relax
\EndOfBibitem
\bibitem[McKenney \emph{et~al.}(2011)McKenney, Simko, Bonham, True, Troyer,
  Hawley, Newcomb, Fazli, Kunju, Nicolas,\emph{et~al.}]{mckenney2011potential}
J.~K. McKenney, J.~Simko, M.~Bonham, L.~D. True, D.~Troyer, S.~Hawley, L.~F.
  Newcomb, L.~Fazli, L.~P. Kunju, M.~M. Nicolas \emph{et~al.}, \emph{The
  Journal of urology}, 2011, \textbf{186}, 465--469\relax
\mciteBstWouldAddEndPuncttrue
\mciteSetBstMidEndSepPunct{\mcitedefaultmidpunct}
{\mcitedefaultendpunct}{\mcitedefaultseppunct}\relax
\EndOfBibitem
\end{mcitethebibliography}
\bibliographystyle{rsc} 

\onecolumn

 \noindent\large{\underline{\textbf{Electronic Supplementary Information}}} \\

\section*{Supplementary Figures}

\renewcommand{\thefigure}{S\arabic{figure}}
\renewcommand{\thetable}{S\arabic{table}}

\begin{figure}[h]
\centering
  \includegraphics[width=17.1cm]{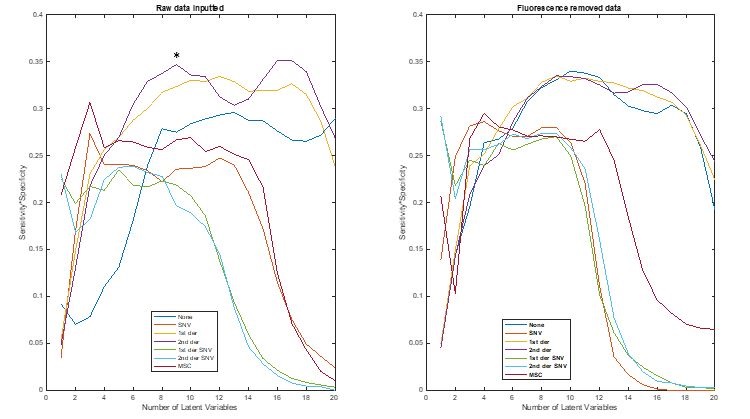}
  \caption{Product of sensitivity and specificity for PLS-DA models constructed on median Raman spectra for non-cancer/cancer classification. Values shown are the mean calculated over 200 random splits of the reference image set. Models were constructed using untreated Raman spectra and six spectral pretreatments: standard normal variate pre-processing (SNV), 1st derivative Savitzky Golay pretreatment (‘1st der’,  window size = 15 points, polynomial order = 3), 2nd derivative Savitzky Golay pretreatment ((‘2nd der’,  window size = 15 points, polynomial order = 3), combinations of SNV followed by 1st or 2nd derivative pretreatment and multiplicative scatter correction (MSC). The optimal PLS-DA model parameters were selected as no pretreatment followed by 2nd derivative preprocessing and a PLS-DA model with 9 latent variables. * 9 LV SG2 model}
  \label{fgr:example}
\end{figure}

\begin{figure}[h]
\centering
  \includegraphics[width=17.1cm]{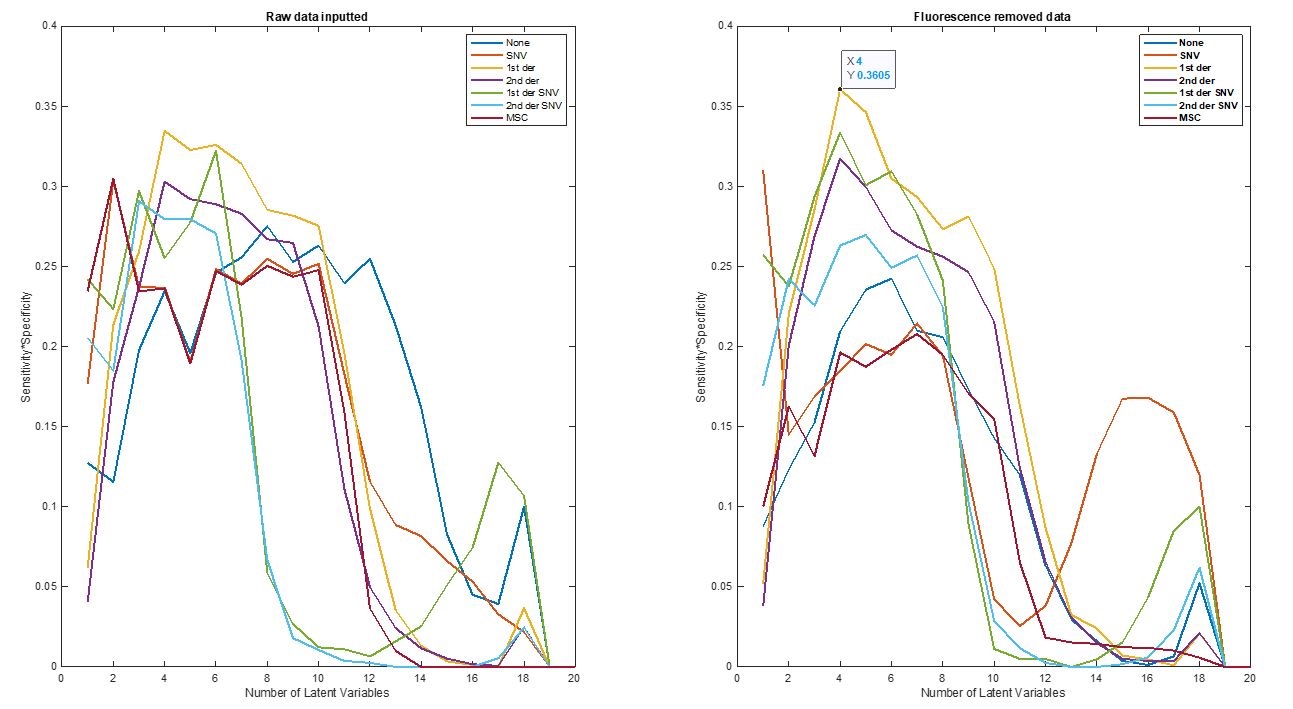}
  \caption{Product of sensitivity and specificity for PLS-DA models constructed on median Raman spectra for Gleason grade 3/ Gleason grade 4. Values shown are the mean calculated over 200 random splits of the reference image set. Models were constructed using untreated Raman spectra and six spectral pretreatments: standard normal variate pre-processing (SNV), 1st derivative Savitzky Golay pretreatment (‘1st der’,  window size = 15 points, polynomial order = 3), 2nd derivative Savitzky Golay pretreatment ((‘2nd der’,  window size = 15 points, polynomial order = 3), combinations of SNV followed by 1st or 2nd derivative pretreatment and multiplicative scatter correction (MSC). The optimal PLS-DA model parameters were selected as fluorescence removal followed by 1st derivative preprocessing and a PLS-DA model with 4 latent variables.}
  \label{fgr:example}
\end{figure}

\clearpage

\section*{Supplementary Tables}

\begin{table}[h]
\small
  \caption{\ Confusion matrix for 9 latent variable PLS-DA model constructed on median Raman spectra (no pretreatment followed by 2nd derivative preprocessing) for Non-Cancer (NC) /Cancer (\text{C}) classification using 5 fold cross validation}
  \label{tbl:example}
  \centering
  \begin{tabular*}{0.48\textwidth}{m{2cm} m{2cm} m{2cm}}
    \hline
     & \textbf{Predicted NC} & \textbf{Predicted C}\\
    \hline
    Actual NC & 32 & 26\\
    Actual C & 27 & 42\\
    \hline
  \end{tabular*}
\end{table}

\begin{table}[h]
\small
  \caption{Confusion matrix for 4 latent variable PLS-DA models constructed on median Raman spectra (pretreated by fluorescence removal followed by 1st derivative preprocessing) for Gleason Grade 3/ Gleason Grade 4 (G3/G4) classification using 5 fold cross validation}
  \label{tbl:example}
  \centering
  \begin{tabular*}{0.48\textwidth}{m{2cm} m{2cm} m{2cm}}
    \hline
     & \textbf{Predicted G3} & \textbf{Predicted G4}\\
    \hline
    Actual G3 & 19 & 20\\
    Actual G4 & 20 & 10\\
    \hline
  \end{tabular*}
\end{table}

\begin{table}[h]
\small
  \caption{Details of non-cancer/cancer (NC/C) and Gleason grade 3/grade 4 (G3/G4) models including diagnosis, modalities and number of
 reference sets. 5-fold cross-validation was used, which incorporated samples from disjoint groups of patients.
}
  \label{tbl:example}
  \centering
  \begin{tabular*}{0.48\textwidth}{m{1.5cm} m{1.7cm} m{2.0cm} m{2.0cm}}
    \hline
    \textbf{Model No.} & \textbf{Diagnosis} & \textbf{Imaging Modality} & \textbf{No. of \newline Reference Sets} \\
    \hline
    1 & NC/C & DP & 1\\
    2 &  & RCI & 1\\
    3 &  & DP+RCI & 1\\    
    4 & G3/G4 & DP & 1\\
    5 &  & RCI & 1\\
    6 &  & DP+RCI & 1\\
    7 &  & DP & 10\\
    8 &  & RCI & 10\\
    9 &  & DP+RCI & 10\\
    \hline
  \end{tabular*}
\end{table}

\begin{table*}[h]
\small
  \caption{Parameters pertaining to optimal SIFT/BoVW/SVM classifiers that used samples from a single randomly selected set of patients as a reference set. Diagnostic tasks are non-cancer/cancer (NC/C) and Gleason grade 3/grade 4 (G3/G4) with digital pathology (DP), Raman Chemical Imaging (RCI) and multimodal (DP+RCI) imaging modalities
}
  \label{tbl:example}
  \centering
  \begin{tabular*}{\textwidth}{@{\extracolsep{\fill}}lllllll}
    \hline
    \textbf{Diagnosis} & \textbf{Imaging Modality} & \textbf{Dictionary Size 1} & \textbf{Dictionary Size 2} & \textbf{C} & \textbf{$\gamma$} & \textbf{Kernel} \\
    \hline
    NC/C & DP & 300 &  & $2^{11}$ & $2^0$ & rbf \\
     & RCI &  & 50 & $2^{8}$ & 2 & rbf \\
     & DP+RCI & 300 & 10 & $2^{12}$ & $2^{-1}$ & rbf \\
     G3/G4 & DP & 200 &  & $2^{7}$ & $2^3$ & rbf \\
     & RCI &  & 5 & $2^{13}$ & $2^3$ & rbf \\
     & DP+RCI  & 300 & 5 & $2^8$ & $2^2$ & rbf \\
    \hline
  \end{tabular*}
\end{table*}

\begin{table*}[h]
\small
  \caption{Parameters pertaining to optimal SIFT/BoVW/SVM classifiers. Each imaging modality was investigated using results from 10 different models, each using different sets of samples for reference images. The diagnostic task is Gleason grade 3/grade 4 (G3/G4) differentiation with digital pathology (DP), Raman Chemical Imaging (RCI) and multimodal (DP+RCI) imaging modalities
}
  \label{tbl:example}
  \centering
  \begin{tabular*}{\textwidth}{@{\extracolsep{\fill}}lllllll}
    \hline
    \textbf{Imaging Modality} & \textbf{Reference Set No.} & \textbf{Dictionary Size 1} & \textbf{Dictionary Size 2} & \textbf{C} & \textbf{$\gamma$} & \textbf{Kernel} \\
    \hline
    DP & 1 & 100 &  & $2^{12}$ & $2^0$ & rbf \\
     & 2 & 500 &  & $2^8$ & $2^1$ & rbf \\
     & 3 & 500 &  & $2^9$ & $2^2$ & rbf \\
     & 4 & 1000 &  & $2^7$ & $2^3$ & rbf \\
     & 5 & 50 &  & $2^{14}$ & $2^{-1}$ & rbf \\
     & 6 & 500 &  & $2^9$ & $2^1$ & rbf \\
     & 7 & 100 &  & $2^{12}$ & $2^0$ & rbf \\
     & 8 & 200 &  & $2^7$ & $2^3$ & rbf \\
     & 9 & 1000 &  & $2^{11}$ &  & linear \\
     & 10 & 500 &  & $2^9$ & $2^1$ & rbf \\
    
    RCI & 1 &  & 10 & $2^{4}$ & $2^3$ & rbf \\
     & 2 &  & 25 & $2^9$ & $2^{-1}$ & rbf \\
     & 3 &  & 5 & $2^{2}$ & $2^2$ & rbf \\
     & 4 &  & 10 & $2^{15}$ & $2^{-1}$ & rbf \\
     & 5 &  & 5 & $2^5$ & $2^{-1}$ & rbf \\
     & 6 &  & 100 & $2^{8}$ & $2^2$ & rbf \\
     & 7 &  & 25 & $2^2$ & $2^2$ & rbf \\
     & 8 &  & 5 & $2^{13}$ & $2^3$ & rbf \\
     & 9 &  & 25 & $2^6$ & $2^1$ & rbf \\
     & 10 &  & 25 & $2^6$ & $2^1$ & rbf \\
    
    DP+RCI & 1 & 50 & 5 & $2^{14}$ & $2^{-1}$ & rbf \\
     & 2 & 200 & 50 & $2^8$ & $2^{1}$ & rbf \\
     & 3 & 100 & 5 & $2^8$ & $2^1$ & rbf \\
     & 4 & 1000 & 10 & $2^{12}$ & $2^{-2}$ & rbf \\
     & 5 & 50 & 5 & $2^{12}$ & $2^{-2}$ & rbf \\
     & 6 & 200 & 5 & $2^{11}$ & $2^{-1}$ & rbf \\
     & 7 & 50 & 50 & $2^8$ & $2^{1}$ & rbf \\
     & 8 & 300 & 5 & $2^8$ & $2^2$ & rbf \\
     & 9 & 1000 & 5 & $2^{9}$ & $2^{2}$ & rbf \\
     & 10 & 75 & 5 & $2^{14}$ & $2^{-3}$ & rbf \\
    
    \hline
  \end{tabular*}
\end{table*}


\balance

\end{document}